\documentclass[aps,twocolumn,prx,amssymb,superscriptaddress,10pt]{revtex4-2}
\usepackage[intlimits]{amsmath}
\usepackage[T1]{fontenc}
\usepackage[utf8]{inputenc}
\usepackage[normalem]{ulem}
\usepackage{dsfont}
\usepackage{comment}
\usepackage{ragged2e}
\usepackage{enumerate}
\usepackage{physics}
\usepackage{color}
\usepackage[dvipsnames]{xcolor}
\usepackage[colorlinks]{hyperref}
\hypersetup{
    colorlinks=true,
    linkcolor=blue,
    filecolor=blue,
    urlcolor=blue,
    citecolor=blue
}
\usepackage{graphicx}
\pdfoutput=1
\graphicspath{{figures/}} 
\usepackage{pgfplots}
\pgfplotsset{compat=1.18} 
\usepackage{tikz}
\usetikzlibrary{arrows.meta}
\tikzset{
myptr/.style={-{Stealth[scale=0.5]}},
}
\usetikzlibrary{patterns}

\def\ee{{\rm e}}
\def\ii{{\rm i}}
\def\kp{\delta}
\newcommand{\mbf}[1]{\mathbf{#1}}
\newcommand{\mbb}[1]{\mathbb{#1}}
\newcommand{\mc}[1]{\mathcal{#1}}
	
\newcommand{\mrm}[1]{\mathrm{#1}}

\newcommand{\ladder}[1]{
\begin{tikzpicture}[scale=#1]
    \def\n{3}\def\a{2}\def\dy{\a*sin(60)}
    \def\linewidth{1.5pt}
    \def\colorone{red!20!green!70!blue!20}
    \def\colortwo{black!90!green!40}
    \def\colorchi{blue!80!green!40}
    \foreach \j in {0,...,\numexpr\n-1\relax} {
        \draw[line width=\linewidth, \colorone] ({\j*\a},0) -- ({(0.5+\j)*\a},{\dy}) -- ({(1+\j)*\a},0);
        }
    \draw[line width=\linewidth, \colorone] ({\n*\a},0) -- ({(0.5+\n)*\a},{\dy});
    \draw[line width=\linewidth, \colortwo, dash pattern={on 6pt off 3pt}] (0,0) -- ({\a*(\n+0.5)},0);
    \draw[line width=\linewidth, \colortwo, dash pattern={on 6pt off 3pt}] (0,{\dy}) -- ({\a*(\n+0.5)},{\dy});
    \foreach \j in {0,...,\numexpr\n\relax} {
        \filldraw[fill=lightgray, draw=Black, line width=\linewidth] ({\a*\j},0) circle (3.5pt);
        \filldraw[fill=lightgray, draw=Black, line width=\linewidth] ({(0.5+\j)*\a},{\dy}) circle (3.5pt);
        }
    \begin{scope}[xshift=3cm, yshift=0.577cm, scale=0.7]
        \draw[->, >=stealth, rounded corners=8pt, line width=1.4pt, \colorchi] (-0.5,-{sqrt(3)/3}) -- (1,-{sqrt(3)/3}) -- (0,{2*sqrt(3)/3}) -- (-0.75,{sqrt(0.25-0.0625)-sqrt(3)/3});
    \end{scope}
    \begin{scope}[xshift=4cm, yshift=1.154cm, scale=0.7, rotate=180]
        \draw[->, >=stealth, rounded corners=8pt, line width=1.4pt, \colorchi] (-0.5,-{sqrt(3)/3}) -- (1,-{sqrt(3)/3}) -- (0,{2*sqrt(3)/3}) -- (-0.75,{sqrt(0.25-0.0625)-sqrt(3)/3}); 
    \end{scope}
    \fill[white] (-0.2,-0.2) rectangle (0.4,{\dy+0.2});
    \fill[white] ({\a*(\n+0.5)-0.4},{\dy+0.2}) rectangle ({0.2+\a*(\n+0.5)},-0.2);
    \node at (2,-.4) {$2n-1$};
    \node at (3,2.2) {$2n$};
    \node at (4,-.4) {$2n+1$};
    \node at (5,2.2) {$2n+2$};
    \node at (2,2.1) {{\color{black!60}$J_2$}};
    \node at (1.8,1.1) {{\color{ForestGreen}$J_1$}};
    \node at (3,{sqrt(3)/3}) {{\color{Blue}$J_\chi$}};
\end{tikzpicture}
}

\newcommand{\parameterizationkappa}[1]{
\begin{tikzpicture}[scale=#1]
    \def\linewidth{1.5pt}
    \def\colorone{red!20!green!70!blue!20}
    \def\colortwo{black!90!green!40}
    \def\colorchi{blue!80!green!40}
    \def\coloraxes{black!70}
    \draw[-, line width=\linewidth, color=\colorone, domain=0:1, samples=81, variable=\k] 
		plot ({\k}, {1-\k});
    \draw[-, line width=\linewidth, color=\colortwo, dashed, domain=0:1, samples=81, variable=\k] 
		plot ({\k}, {\k/2});
    \draw[-, line width=\linewidth, color=\colorchi, domain=0:1, samples=81, variable=\k] 
		plot ({\k}, {sqrt(\k)*sqrt(1-\k)});
    \draw[<->, >=stealth, line width=\linewidth, color=\coloraxes] (1.3,0) -- (0,0) -- (0,1.3);
    \draw[-, line width=1, \coloraxes] (1,-0.1) -- (1,0.1);
    \draw[-, line width=1, \coloraxes] (-0.1,1) -- (0.1,1);
    \node at (0.5, 1.6) {Couplings:};
    \node at (0.5,-0.25) {$\eta$};
    \node at (1,-0.22) {\scriptsize$1$};
    \node at (-0.22,1) {\scriptsize$1$};
\end{tikzpicture}
}

\newcommand{\circuit}{%
\begin{tikzpicture}[baseline=(current  bounding  box.center),scale=1]
    \foreach \x in {-5,...,5}
    {
    \draw[black, line width=0.375mm] (1+0.5*\x,-1.75) -- (1+0.5*\x,2.25);
    }
    
    \draw[white!30!red, line width=0.625mm] (0.75,0) -- (0.75,0.5);
    \draw[white!30!red, line width=0.625mm] (0.75,0.5) to[out=90, in=180] (0.875,0.625);
    \draw[white!30!red, line width=0.625mm] (0.75,0) to[out=-90, in=0] (0.625,-0.125);
    \draw[white!30!red, line width=0.625mm] (0.75-1,0) -- (0.75-1,0.5);
    \draw[white!30!red, line width=0.625mm] (0.75-1,0.5) to[out=90, in=0] (0.625-1,0.625);
    \draw[white!30!red, line width=0.625mm] (0.75-1,0) to[out=-90, in=180] (0.875-1,-0.125);
    \draw[white!30!red, line width=0.625mm] (0.875-1,-0.125) -- (0.625,-0.125);
    \draw[white!30!red, line width=0.625mm] (0.75,1.5) -- (0.75,2);
    \draw[white!30!red, line width=0.625mm] (0.75,1.5) to[out=-90, in=180] (0.875,1.375);
    \draw[white!30!red, line width=0.625mm] (0.75,2) to[out=90, in=0] (0.625,2.125);
    \draw[white!30!red, line width=0.625mm] (0.75-1,1.5) -- (0.75-1,2);
    \draw[white!30!red, line width=0.625mm] (0.75-1,1.5) to[out=-90, in=0] (0.625-1,1.375);
    \draw[white!30!red, line width=0.625mm] (0.75-1,2) to[out=90, in=180] (0.875-1,2.125);
    \draw[white!30!red, line width=0.625mm] (0.875-1,2.125) -- (0.625,2.125);
    \draw[white!30!red, line width=0.625mm] (0.625-1,1.375) -- (0.375-1,1.375);
    \draw[white!30!red, line width=0.625mm] (0.375-1,1.375) to[out=180, in=90] (0.25-1,1.25);
    \draw[white!30!red, line width=0.625mm] (0.625-1,0.625) -- (0.375-1,0.625);
    \draw[white!30!red, line width=0.625mm] (0.375-1,0.625) to[out=180, in=-90] (0.25-1,0.75);
    \draw[white!30!red, line width=0.625mm] (0.25-1,0.75) -- (0.25-1,1.25);
    \draw[white!30!red, line width=0.625mm] (0.875,1.375) -- (1.125,1.375);
    \draw[white!30!red, line width=0.625mm] (0.875,1.375) to[out=0, in=90] (1.25,1.25);
    \draw[white!30!red, line width=0.625mm] (0.875,0.625) -- (1.125,0.625);
    \draw[white!30!red, line width=0.625mm] (1.125,0.625) to[out=0, in=-90] (1.25,0.75);
    \draw[white!30!red, line width=0.625mm] (1.25,0.75) -- (1.25,1.25);

    \foreach \x in {-1,...,4}
        {
        \draw[fill=white, opacity=1, rounded corners = 2,thick] (-0.625+\x,0.75) rectangle ++(0.75,0.5);
        \draw[fill=red!20!green!70!blue!20, opacity=1, rounded corners = 2,thick] (-0.625+\x,0.75) rectangle ++(0.75,0.5);
        \draw[fill=white, opacity=1, rounded corners = 2,thick] (-0.625+\x,-0.75) rectangle ++(0.75,0.5);
        \draw[fill=red!20!green!70!blue!20, opacity=1, rounded corners = 2,thick] (-0.625+\x,-0.75) rectangle ++(0.75,0.5);
        }
    \foreach \x in {-1,...,1}
        {
        \draw[fill=white, opacity=1, rounded corners = 2,thick] (-0.125+2*\x,1.5) rectangle ++(0.75,0.5);
        \draw[pattern=crosshatch, pattern color=blue!80!green!40, opacity=1, rounded corners = 2,thick] (-0.125+2*\x,1.5) rectangle ++(0.75,0.5);
        \draw[fill=white, opacity=1, rounded corners = 2,thick] (-0.125+2*\x,0) rectangle ++(0.75,0.5);
        \draw[pattern=north west lines, pattern color=black!90!green!40, opacity=1, rounded corners = 2,thick] (-0.125+2*\x,0) rectangle ++(0.75,0.5);
        \draw[fill=white, opacity=1, rounded corners = 2,thick] (0.875+2*\x,0) rectangle ++(0.75,0.5);
        \draw[fill=white, opacity=1, rounded corners = 2,thick] (0.875+2*\x,0) rectangle ++(0.75,0.5);
        \draw[pattern=crosshatch, pattern color=blue!80!green!40, opacity=1, rounded corners = 2,thick] (0.875+2*\x,0) rectangle ++(0.75,0.5);
        \draw[fill=white, opacity=1, rounded corners = 2,thick] (0.875+2*\x,-1.5) rectangle ++(0.75,0.5);
        \draw[pattern=north west lines, pattern color=black!90!green!40, opacity=1, rounded corners = 2,thick] (0.875+2*\x,-1.5) rectangle ++(0.75,0.5);
        }

    \fill[white] (-2.75,-1.8) rectangle ++(1,4);
    \fill[white] (3.75,-1.8) rectangle ++(1,4);

    \draw[-{Stealth[scale=0.75]}, black, line width=0.5mm] (0.25-2.125,-1.875) -- (0.25-2.125,-0.875);
    \draw[-{Stealth[scale=0.75]}, black, line width=0.5mm] (0.25-2.125,-1.875) -- (0.25-1.125,-1.875);
    \node[anchor=east] at (0.25-2.125,-0.875){$t$};
    \node[anchor=north] at (0.25-1.125,-1.875){$x$};
    \node[anchor=south] at (0.25,2.25){$U_{[4j-1,4j+2]}$};

    \draw[black,thick] (-2+6.325,0.875) -- (-2+6.325,1.5);
    \draw[black,thick] (-2+6.675,0.875) -- (-2+6.675,1.5);
    \draw[fill=red!20!green!70!blue!20, opacity=1, rounded corners = 2,thick] (-2+6.25,1) rectangle ++(0.5,0.375);
    \draw[black,thick] (-2+6.325,0.125) -- (-2+6.325,0.75);
    \draw[black,thick] (-2+6.675,0.125) -- (-2+6.675,0.75);
    \draw[fill=white, opacity=1, rounded corners = 2,thick] (-2+6.25,0.25) rectangle ++(0.5,0.375);
    \draw[pattern=crosshatch, pattern color=blue!80!green!40, opacity=1, rounded corners = 2,thick] (-2+6.25,0.25) rectangle ++(0.5,0.375);
    \draw[black,thick] (-2+6.325,-0.625) -- (-2+6.325,0);
    \draw[black,thick] (-2+6.675,-0.625) -- (-2+6.675,0);
    \draw[fill=white, opacity=1, rounded corners = 2,thick] (-2+6.25,-0.5) rectangle ++(0.5,0.375);
    \draw[pattern=north west lines, pattern color=black!90!green!20, opacity=1, rounded corners = 2,thick] (-2+6.25,-0.5) rectangle ++(0.5,0.375);
    \node[anchor=west] at (-2.25+7,1.175){$\check{R}(\tau)$};
    \node[anchor=west] at (-2.25+7,0.425){$\check{R}(\tau\!-\!\delta)$};
    \node[anchor=west] at (-2.25+7,-0.325){$\check{R}(\tau\!+\!\delta)$};
    \node[anchor=west] at(-2+6,1.875){Gates:};

\end{tikzpicture}
}

\begin{document}
\makeatletter
\renewcommand\@makefnmark{}
\renewcommand\@makefntext[1]{\noindent #1}
\makeatother

\title{Superdiffusion and anomalous fluctuations in chiral integrable dynamics}

\author{Cristiano Muzzi }
\email{cmuzzi@sissa.it}
\affiliation{%
International School for Advanced Studies (SISSA), Via Bonomea 265, Trieste I-34136, Italy
}
\affiliation{INFN, Sezione di Trieste, via Valerio 2, 34127 Trieste, Italy}
\author{Devendra Singh Bhakuni}
\email{dbhakuni@ictp.it}
\affiliation{%
The Abdus Salam International Centre for Theoretical Physics (ICTP),
Strada Costiera 11, Trieste I-34151, Italy
}
\author{Marcello Dalmonte}
\email{mdalmont@ictp.it}
\affiliation{%
The Abdus Salam International Centre for Theoretical Physics (ICTP),
Strada Costiera 11, Trieste I-34151, Italy
}
\author{Lenart Zadnik}
\email{lenart.zadnik@fmf.uni-lj.si}
\affiliation{%
Faculty of Mathematics and Physics, University of Ljubljana, Jadranska 19, SI-1000 Ljubljana, Slovenia
}
\author{Hernan B. Xavier}
\email{hxavier@ictp.it}
\thanks{\\$*, \dag$ These authors contributed equally to this work}

\affiliation{%
International School for Advanced Studies (SISSA), Via Bonomea 265, Trieste I-34136, Italy
}
\affiliation{%
The Abdus Salam International Centre for Theoretical Physics (ICTP),
Strada Costiera 11, Trieste I-34151, Italy
}

\begin{abstract}
Symmetries strongly influence transport properties of quantum many-body systems, and can lead to deviations from the generic case of diffusion. In this work, we study the impact of time-reversal symmetry breaking on the transport and its universal aspects in integrable chiral spin ladders. We observe that the infinite-temperature spin transport is superdiffusive with a dynamical critical exponent $z=3/2$ matching the one of the Kardar-Parisi-Zhang (KPZ) universality class, which also lacks the time reversal symmetry. However, we find that fluctuations of the net magnetization transfer deviate from the KPZ predictions. Moreover, the full probability distribution of the associated spin current obeys fluctuation symmetry despite broken time-reversal and space-reflection symmetries. To further investigate the role of conserved quantities, we introduce an integrable quantum circuit that shares the essential symmetries with the chiral ladder, and which exhibits analogous dynamical behaviour in the absence of energy conservation. Our work shows that time-reversal symmetry breaking is compatible with superdiffusion, but insufficient to stabilize the KPZ universality in integrable systems. This suggests that additional fundamental features are missing in order to identify the emergence of such dynamics in quantum matter. 
\end{abstract}

\maketitle

\section{Introduction}

Understanding charge transport is one of the archetypal problems in the physics of strongly correlated quantum matter. It has been addressed in various experimentally relevant contexts, ranging from quark dynamics in heavy-ion collisions~\cite{Romatschke2007Viscosity,Teaney2010Viscous,Snellings2011Elliptic,Busza2018Heavy}
to correlated electrons and synthetic quantum matter~\cite{Schneider2012Fermionic,Trotzky2012Probing,Fukuhara2014Far-from-equilibrium,Georgescu2014Quantum,Brown2015Two-dimensional,Chien2015Quantum,Gross2017Quantum,Nichols2019Spin,Brown2019Bad}. Recently, there has been a growing interest in understanding exceptions to the generic case of diffusive charge dynamics~\cite{ZnidarichSpin,ljubotina2019kardar,DupontMoore,barlev2015absence,agarwal2015anomalous,Bar_Lev_2017,feldmeier2020anomalous,morningstar2020kinetically,moudgalya2021spectral,gromov2020fracton,iaconis2021multipole,znidaric2024superdiffusive,ljubotina2023superdiffusive,yupeng2024superdiffusive,bhakuni2025anomalously,moca2025dynamicscalingfamilyvicsekuniversality}. A notable one arises in integrable spin chains, where extensive theoretical~\cite{ZnidarichSpin,ljubotina2019kardar,DupontMoore} and experimental~\cite{scheie2021detection,wei2022quantum,rosenberg2024dynamics} evidence suggests the existence of superdiffusive magnetization transport at infinite temperature~\cite{Bulchandani_2021,Gopalakrishnan_hot,ilievski2018,ilievski2021superuniversality,GopalakrishnanKineticTheory}. In particular, spin superdiffusion has been established in systems exhibiting both integrability and continuous non-Abelian symmetries~\cite{ilievski2021superuniversality}, as reported, e.g., in the case of the isotropic Heisenberg model~\cite{ZnidarichSpin,TakeuchiPartialKPZ} or its integrable semiclassical (large-spin) limit---the Landau-Lifshitz model~\cite{DasLandauLifshitz,KrajnikLandauLifshitz}. 

However, the impact of other fundamental symmetries, such as the time-reversal and the space-reflection ones, on superdiffusive transport has largely remained obscure. This lack of understanding is particularly impactful, since one of the most discussed emergent transport universality classes is the Kardar-Parisi-Zhang one, which---contrary to all spin models studied so far---posits the \emph{absence} of time reversal invariance~\cite{NonlinearFH_JDN}. There is, so far, no evidence that superdiffusion can be robust to time-reversal symmetry breaking, and if so, what changes in the universal behaviour can this breaking induce. 

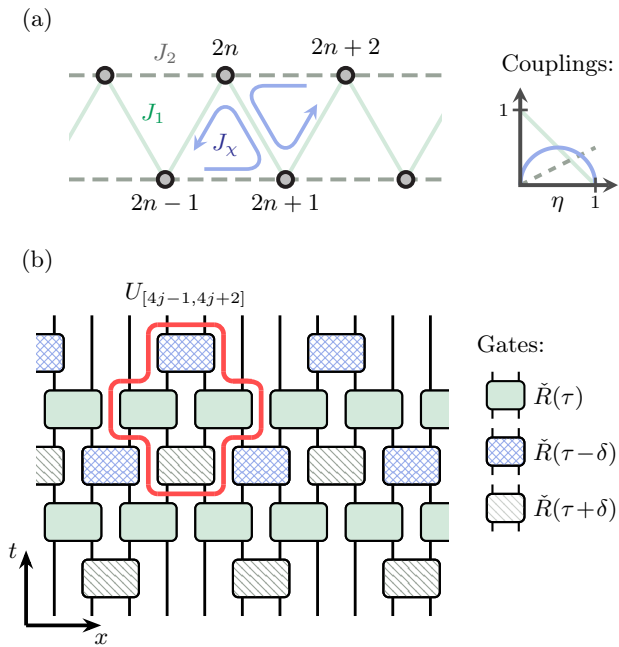
\begin{figure}[t]
    \centering
    \hspace{-2em}
    \begin{tikzpicture}
        \node (A) at (-0.5,0) {\ladder{.8}};
        \node at (3.5,-0.1) {\parameterizationkappa{1}};
        \node (B) at (0,-4.5) {\circuit};
        \node at (-3.4,1.4) {(a)};
        \node at (-3.4,-1.8) {(b)};
    \end{tikzpicture}
    \caption{\textbf{SU(2)-invariant spin-1/2 systems considered in this work:}
    (a) Triangular ladder parametrized by $\eta\in[0,1]$, with nearest-neighbour exchange $J_1=1-\eta$, next-nearest-neighbour exchange $J_2=\eta/2$, and uniform three-spin chiral interaction $J_\chi=\sqrt{\eta(1-\eta)}$.
    (b) Brickwork circuit constructed from the composition of four elementary gates $\check{R}(\tau)=(\mathds{1}+\ii\tau P)/(1+\ii\tau)$, some of which are deformed by the parameter $\delta=-\sqrt{\eta/(1-\eta)}$. Here, $P$ is the permutation of two neighbouring spins $1/2$. The composite four-site unitary gate $U_{[4j-1,4j+2]}$, acting in sites $4j\!-\!1,4j,4j\!+\!1,4j\!+\!2$, is encircled in red. Up to $O(\tau^3)$ corrections, it is equivalent to the exponential of the four-site Hamiltonian density of the chiral spin ladder shown in panel (a). The continuous- and discrete-time dynamics are equivalent in the limit $\tau\to0$.}
    \label{fig:ladder}
\end{figure}

A related question that has not been adequately addressed is how the space-time symmetries affect the fluctuations of charge and its associated current. While the time-reversal symmetry implies that the ballistic-scale current fluctuations in maximum-entropy states have no preferred direction~\cite{doyon-perfetto-sasamoto-yoshimura-2023}, the effect of broken time reversal symmetry on the fluctuations remains less clear. In particular, there exist dynamical systems in which the fluctuation symmetry seems to be absent not due to the lack of time-reversal invariance, but rather due to the particular way in which the space-reflection symmetry is broken~\cite{Ratchets}. What is the precise relation between the space-time symmetries and the symmetries of charge and current fluctuations is a particularly timely question, especially in light of the recent theoretical and experimental studies, which have reported discrepancies with the KPZ predictions concerning the higher-order fluctuations of magnetization transfer in the Heisenberg spin chain~\cite{rosenberg2024dynamics,valli2024efficientcomputationcumulantevolution}. 

Here, we report the observation of superdiffusive spin transport at infinite temperature in a class of SU(2)-invariant integrable spin models where time-reversal symmetry breaking can be controlled. Specifically, we study both a continuous-time dynamics in an integrable chiral spin ladder and an integrable (discrete-time) quantum circuit that shares the same key symmetries, allowing us to also explore the effects of broken energy conservation. In both cases we show that, even in the absence of time-reversal symmetry, the spin transport is superdiffusive with a dynamical exponent $z=3/2$, but with magnetization-transfer statistics that is incompatible with the KPZ universality class. Moreover, we observe that the associated spin current probability distribution obeys a fluctuation symmetry. Surprisingly, the latter is not a consequence of time-reversal and space-reflection symmetries, since they are both broken. Instead, the fluctuation symmetry follows from another spatial $C_2$ symmetry (invariance under $\pi$-rotations of the ladder).

The platform for our study is a chiral spin ladder~\cite{Popkov1993Antichiral,Frahm1996Integrable,Frahm1997Properties}, depicted in Fig.~\hyperref[fig:ladder]{1(a)}, whose continuous-time dynamics and symmetries are described in Section~\ref{sec:chiral-dynamics}. To explore features of the charge transport, we employ two complementary approaches. In Section~\ref{sec:hydrodynamic-approach} we first determine the transport regime by considering the dynamical structure factor. This approach involves the use of thermodynamic Bethe ansatz combined with hydrodynamics~\cite{doyon2017drude,doyon2020lecture,Doyon_2025}.
In Section~\ref{sec:gen_func} we next identify the transport universality class by considering the full counting statistics of the magnetization transfer. To access it, we employ the recently developed quantum generating function approach, which is based on tensor-network methods~\cite{valli2024efficientcomputationcumulantevolution}. Finally, in Section~\ref{sec:circuit} we relax the assumption of energy conservation and study transport in an integrable quantum circuit that has the same essential symmetries as the chiral spin ladder, and which fits the contemporary digital quantum simulation platforms. The brickwork-type circuit is depicted in Fig.~\hyperref[fig:ladder]{1(b)}. The key symmetries of the chiral spin ladder are retained by choosing the unitary gates which, up to the second order in the time step, match the exponential of the spin ladder's Hamiltonian density. We summarize our findings in Section~\ref{sec:conclusions}.

\section{Continuous-time chiral dynamics}
\label{sec:chiral-dynamics}

We consider a triangular two-leg spin ladder, with sites labelled by indices $\{1,\dots,L\}$ and arranged in a zigzag geometry, as shown in Fig.~\hyperref[fig:ladder]{1(a)}. The system dynamics is governed by the Hamiltonian~\cite{Frahm1997Properties}
\begin{align}\label{eq:hamiltonian}
    &H= H_0+H_\chi,\nonumber\\
    &H_0=\sum^{L}_{n=1}
    J_{1}\mbf{S}_{n}\cdot\mbf{S}_{n+1} 
    +J_{2}\mbf{S}_{n}\cdot\mbf{S}_{n+2},\nonumber\\
    &H_\chi=J_{\chi}\sum^{L}_{n=1}(-1)^n\mbf{S}_{n}\cdot(\mbf{S}_{n+1}\times\mbf{S}_{n+2}),
\end{align}
where $\mbf{S}_n$ is the vector of spin-$1/2$ operators on site $n$, with components $S^\alpha_n=\frac{1}{2}\sigma_{n}^\alpha$ for $\alpha\in\{x,y,z\}$.
The first term, $H_0$, describes a frustrated Heisenberg chain with nearest- and next-nearest-neighbour couplings $J_1$ and $J_2$, respectively. The second term, $H_\chi$, introduces a chiral interaction of strength $J_\chi$~\footnote{%
Note that our definition of $H_\chi$ differs by a factor of $1/2$ from Ref.~\cite{Frahm1997Properties}, but is consistent with the model Hamiltonian reported in Ref.~\cite{Frahm1996Integrable}, and Bethe ansatz equations found in Refs.~\cite{Popkov1993Antichiral,Frahm1996Integrable,Frahm1997Properties}.
}. The staggered factor $(-1)^n$ accounts for the alternating site orientations in the plaquettes, and ensures a uniform spin-circulation sense across the ladder.

The ladder preserves a global SU(2) spin symmetry for all parameter values. 
However, the chiral coupling $J_\chi$ reduces the translational symmetry from one-site to two-site periodicity.
It also breaks time-reversal symmetry, $\mc{T}$, and the site parity, $\mc{P}_{s}$, while preserving their combined action $\mathcal{P}_{s}\mathcal{T}$. The ladder also retains the link parity $\mathcal{P}_{\ell}$, corresponding to a $C_2$ rotation about the central link. 
We note, in passing, that there is a significant interest in the low-energy properties of such chiral spin systems~\cite{WenPSG}, 
as they provide a valuable platform for investigating the stability of chiral spin-liquid phases
~\cite{Gorohovsky2015Chiral,willsher2025dynamics,oliviero2022noncoplanar}.

In the following, we focus on the integrable line
\begin{align}\label{eq:integrable_line}
    J_1=1-\eta,\quad
    J_2=\frac{\eta}{2},\quad
    J_\chi=\sqrt{\eta(1-\eta)},
\end{align}
with $\eta \in [0,1]$, where the model admits an exact Bethe-ansatz solution~\cite{Popkov1993Antichiral,Frahm1996Integrable,Frahm1997Properties}.As further illustrated in Fig.~\hyperref[fig:ladder]{1(a)}, by varying $\eta$ continuously, one interpolates between a single Heisenberg model (at $\eta=0$), and two decoupled chains (at $\eta=1$). The chiral coupling $J_\chi$ vanishes at both ending points, and reaches its maximum at $\eta=1/2$.

\section{Dynamical structure factor}
\label{sec:hydrodynamic-approach}

To investigate charge transport, we will first consider the moments $m_n(t):= \int\!{\rm d}x \, x^n S(x,t)$ of the dynamical structure factor
\begin{align}
    S(x,t)= \langle q(x,t) q(0,0)\rangle- \langle q(x,t)  \rangle \langle q(0,0)\rangle.
\end{align}
Here, $q(x,t)$ is the charge density at coarse grained coordinates $x$ and $t$, and $\langle \bullet\rangle$ denotes the expectation value in the thermodynamic state of interest. Transport regime is characterized by the dynamical exponent $z$, identified from the scaling of the second moment,
\begin{align}
    m_2(t)\sim t^{2/z}.
    \label{eq:dynamical-exponent}
\end{align}
Depending on the value of $z$, we distinguish ballistic ($z=1$), superdiffusive ($1<z<2$), diffusive ($z=2$), and subdiffusive ($z>2$) transport.  

In integrable spin models, such as the chiral ladder that we consider herein, the structure factor can be decomposed in terms of ballistically spreading quasiparticles---magnons and their bound states---with rapidity $\lambda$ and total spin $s$~\cite{doyon2017drude,doyon2020lecture}. On ballistic scale (large $x$ and $t$, with the ratio $x/t$ fixed) such a decomposition reads
\begin{align}
\label{eq:struc-fact-hydro}
    S(x,t)\simeq
    \sum_{s=1}^\infty\!\int \!\!{\rm d}\lambda \,\delta\!\left(x\!-\!v_s t\right)\!\rho^{\rm t}_s n_s(1\!-\!n_s)\left(q^{\rm dr}_s\right)^2,
\end{align}
where we have suppressed the dependence on $\lambda$. The total state density $\rho^{\rm t}_s(\lambda)$ and the occupancy ratio $n_s(\lambda)$ determine the thermodynamic state $\langle\bullet\rangle$. Charge $q^{\rm dr}_s(\lambda)$ is dressed due to scattering between the quasiparticles which move ballistically with effective velocities $v_s(\lambda)$. All of the above ingredients are properties of the thermodynamic state and are obtained from the thermodynamic Bethe ansatz (TBA). In particular, we will consider spin transport at infinite temperature and chemical potential $\mu$, i.e., in the grand-canonical state
\begin{align}
    \langle\bullet\rangle={\rm Tr}\left[(\bullet)\varrho_\mu\right],\qquad \varrho_\mu = \frac{\ee^{-\mu \sum_j S_j^z}}{{\rm Tr}\left(\ee^{-\mu \sum_j S_j^z}\right)},
    \label{eq:grand-canonical-state}
\end{align} 
in which TBA and hydrodynamics can be most easily applied.

The Hamiltonian of the integrable chiral spin ladder turns out to be a sum of the first two conserved quantities in the integrable Trotterization of the Heisenberg model~\cite{vanicat2018} (see Appendix~\ref{app:transfer-matrices} for details). The ingredients for the hydrodynamic decomposition of the spin structure factor in Eq.~\eqref{eq:struc-fact-hydro} therefore closely resemble those derived in Refs.~\cite{Ratchets,paletta2025}, in the context of integrable circuits (i.e., integrable Floquet systems). At infinite temperature the occupancy ratio is~\cite{takahashi1999}
\begin{align}
    n_s=\frac{1}{\chi_s^2},\qquad \chi_s:=\frac{\sinh[(s+1)\frac{\mu}{2}]}{\sinh(\frac{\mu}{2})}.
    \label{eq:occupancy-ratio}
\end{align}
The relation between the Hamiltonian in Eq.~\eqref{eq:hamiltonian} and the integrable Trotterization of the Heisenberg model moreover allows us to express the total state density as
\begin{align}
    &\rho^{\rm t}_s(\lambda)=\tilde{\rho}^{\rm t}_s(\lambda)+\tilde{\rho}^{\rm t}_s\left(\lambda-\delta\,\right),\quad \delta=-\sqrt{\frac{\eta}{1-\eta}},\notag\\
    &\tilde{\rho}^{\rm t}_s(\lambda)=\frac{\chi_s}{2\pi\chi_1 }\left(\frac{\tilde{p}_s'(\lambda)}{\chi_{s-1}}-\frac{\tilde{p}_{s+2}'(\lambda)}{\chi_{s+1}}\right).
    \label{eq:state-density}
\end{align}
Here, $(\bullet)'$ is the rapidity derivative, while $\tilde{\rho}^{\rm t}_s(\lambda)$ and $\tilde{p}_s(\lambda)=\frac{\ii}{2}\log[(\lambda+\ii \frac{s}{2})/(\lambda-\ii \frac{s}{2})]$ are the total state density and the bare quasiparticle momentum in the Heisenberg model, respectively. Finally, the effective quasiparticle velocity and the dressed charge (i.e., magnetization, dressed due to the inter-particle scattering processes) are obtained as
\begin{align}
    v_s(\lambda)\!=\!-\frac{1}{2}\partial_\lambda\!\log\rho^{\rm t}_s\!(
    \lambda),
    \label{eq:velocity}
\end{align}
and
\begin{align}
    q_s^{\rm dr}\!=\!\partial_\mu\!\log(n_s^{-1}\!-\!1),
    \label{eq:charge}
\end{align}
respectively. We relegate the derivations of the above expressions and technical details to Appendix~\ref{app:TBA}.

\subsection{Drude weights and ballistic transport}
\label{sec:drude-ballistic}

Due to the relation to the Heisenberg circuit [see, e.g., Eq.~\eqref{eq:state-density}], we expect the chiral spin ladder to support ballistic spin transport away from half filling, i.e., for $\mu>0$. In a system with ballistic charge transport ($z=1$), the second moment of the structure factor scales as $m_2(t)\simeq D t^2$ at long times. Here, the finite coefficient $D>0$ is the charge Drude weight, which can be expressed as
\begin{align}
    D&=\lim_{t\to\infty}\int\dd x \, \left(\frac{x}{t}\right)^2 S(x,t)\notag\\
    &=\sum_s\int{\rm d}\lambda \,\rho^{\rm t}_s v_s^2\, n_s(1-n_s)\left(q^{\rm dr}_s\right)^2,
    \label{eq:drude-weight}
\end{align}
where the hydrodynamic resolution in Eq.~\eqref{eq:struc-fact-hydro} has been applied. Another quantity of interest in ballistic systems is the first absolute moment of $S(x,t)$, which defines the so-called Drude self-weight through the scaling $\int{\rm d}x |x|S(x,t)\simeq D_{\rm sw}t$. The hydrodynamic resolution of the Drude self-weight is then
\begin{align}
    D_{\rm sw}=\sum_s\!\int\! {\rm d}\lambda \,\rho^{\rm t}_s |v_s|\, n_s(1-n_s)\left(q^{\rm dr}_s\right)^2.    
    \label{eq:drude-self-weight}
\end{align}
Since the first absolute moment of $S(x,t)$ can also be obtained from the QGF method, described in Section~\ref{sec:gen_func}, it will serve as a useful benchmark. 

For the spin transport, $Q=S^z_{\rm total}=\sum_{j=1}^L S^z_j$, we find both the Drude weight and the Drude self-weight to be finite away from half-filling ($\mu\ne 0$), consistently with ballistic spin transport. They are computed by truncating the sum in Eq.~\eqref{eq:drude-self-weight} at some $s_{\rm max}$ and numerically integrating over $\lambda$. We obtain, for example, $D_{\rm sw}(\eta\!=\!0,\mu\!=\!0.5)\approx0.0482(1)$ and $D_{\rm sw}(\eta\!=\!0.5,\mu\!=\!0.5)\approx 0.0412(4)$. These results are benchmarked against the QGF method, described in Section~\ref{sec:gen_func}.

\subsection{Superdiffusion from ballistically propagating quasiparticles}
\label{sec:superdiffusion-self-consistent}

Having established ballistic transport for $\mu\neq 0$, we now argue the presence of superdiffusion at half-filling ($\mu=0$), following the arguments in Refs.~\cite{ilievski2021superuniversality, GopalakrishnanKineticTheory}. For small $\mu$, the dressed charge in Eq.~\eqref{eq:charge} reads
\begin{align}
    q^{\rm dr}_{s}=\frac{1}{6}(s+1)^2 \mu+ O(\mu^3).
    \label{eq:charge-expansion}
\end{align}
Since it vanishes as $\mu\to 0$, so does the Drude weight in Eq.~\eqref{eq:drude-weight}, and the spin transport at half-filling is sub-ballistic. To determine the dynamical exponent $z$, we revisit the variance of the dynamical structure factor, assuming the hydrodynamic form
\begin{align}
    m_2(t)\simeq t^2 \sum_s \int\dd\lambda \,\rho^{\rm t}_s v_s^2\, n_s(1-n_s)\left(q^{\rm dr}_s\right)^2,
    \label{eq:second-moment-hydro}
\end{align}
and consider the dressed charge as a fluctuating quantity. Specifically, charge is carried by quasiparticles moving ballistically through a background that instead evolves sub-ballistically. Quasiparticles experience fluctuations in the magnetic field $h\sim\mu$, which directly translate into corresponding fluctuations of the dressed charge, according to Eq.~\eqref{eq:charge-expansion}. For sufficiently fast quasiparticles, the background can be treated as frozen. When such quasiparticles traverse the distance $|v_s t|$, they experience magnetic-field fluctuations that scale as $\langle h^2\rangle\sim |v_s t|^{-1}$~\footnote{A quasiparticle traversing the distance $\ell\sim|v_s t|$ encounters $\ell$ spins that can be thought of as randomly chosen from some probability distribution with a finite variance $\sigma^2$ and zero mean. The magnetic field felt by the quasiparticle corresponds to the mean field $h$ of the sequence of $\ell$ traversed spins. For large $\ell$ (or $t$), by the central limit theorem, $\sqrt{\ell}h$ is normally distributed with zero mean and variance $\sigma^2$. Therefore, $\langle h^2\rangle\sim |v_s t|^{-1}$.}. The square of the dressed charge in Eq.~\eqref{eq:charge-expansion} then scales as $\left(q^{\rm dr}_s\right)^2\sim (s+1)^4/|v_s t|$. When $\mu\to 0$, the effective velocity in Eq.~\eqref{eq:velocity} scales as $|v_s(\lambda)|\sim s^{-1}$. For the occupancy ratio in Eq.~\eqref{eq:occupancy-ratio}, we find $n_s\sim (s+1)^{-2}$, except for heavy (slow) quasiparticles with $s>s_*\sim h^{-1}$, where $s_*$ serves as a cutoff. For these heavy quasiparticles, $n_s$ is instead exponentially suppressed in $s$. 

Using these scaling forms in Eq.~\eqref{eq:second-moment-hydro}, we obtain
\begin{align}
    m_2(t)\sim t \sum_{s=1}^{s_*} s(s+2) I_s,
\end{align}
where we have denoted $I_s=\int \dd\lambda\,\rho^{\rm t}_s |v_s|$. This integral scales as $I_s\sim s^{-2}$, as long as $s < s_*\sim h^{-1}$. We thus arrive at $m_2(t)\sim t s_*$, however, the cutoff $s_*\sim h^{-1}$ is itself a function of time. Indeed, the effective field that arises from the background fluctuations, and which is felt by quasiparticles with $s\sim s_*$, scales as $h\sim |v_{s_*}t|^{-1/2}\sim |t/s_*|^{-1/2}$. From here, we find $s_*(t)\sim t^{1/3}$ and $m_2(t)\sim t^{4/3}$~\footnote{As an additional check of self-consistency, we note that for slow quasiparticles with $s\sim s_*$ the background cannot be treated as frozen. These quasiparticles move with the background, i.e., they diffuse with a time-dependent diffusion constant $s_*(t)$, since $m_2(t)\sim s_*(t) t$. Relating the length scales as $|v_{s_*}t|^2\sim s_* t$, where $|v_{s_*}|\sim s_*^{-1}$, we again find $s_*(t)\sim t^{1/3}$.}.
According to Eq.~\eqref{eq:dynamical-exponent}, the transport is thus superdiffusive, with dynamical exponent $z=3/2$. 

\section{Full-counting statistics of the charge transfer}
\label{sec:gen_func}

A complementary way to characterize transport is through the full counting statistics of charge transfer, defined by the generating function 
\begin{align}
    G(\lambda,t)=\ev*{\ee^{\ii\lambda\Delta Q_\mc{S}(t)}},\
    \Delta Q_\mc{S}(t)=Q_\mc{S}(t)-Q_\mc{S}(0),
\end{align}
where $Q_\mc{S}(t)\equiv\int_{\mathcal{S}} {\rm d}x \,q(x,t)$ denotes the total charge in region $\mathcal{S}$ at time $t$.
Derivatives of the generating function offer access to the moments of the net charge transfer, defined as $\mu_n(t):=\ev{[\Delta Q_\mc{S}(t)]^n}$.
In particular, derivatives of $\log G(\lambda,t)$ yield the corresponding cumulants $\kappa_n(t)$, which isolate the connected components of the charge–transfer fluctuations.
The moments $\mu_n(t)$ encode transport properties and can be used to identify the corresponding universality class. By virtue of the continuity equation, the charge-transfer moments are related to the moments of the time-integrated currents flowing out of the region $\mathcal{S}$.

Similarly to the second moment of the structure factor, $m_2(t)$, the dynamical exponent $z$ also governs the growth of the second moment of the transferred charge, $\mu_2(t)\sim t^{1/z}$. In fact, choosing the region $\mathcal{S}$ as the left-half of the system, $\mu_2(t)$ coincides with the first absolute moment of the structure factor $S(x,t)$. In ballistic systems, we therefore have $\mu_2(t)\simeq D_{\rm sw}t$. 

\begin{figure*}[ht!]
    \centering
    \includegraphics[width=\linewidth]{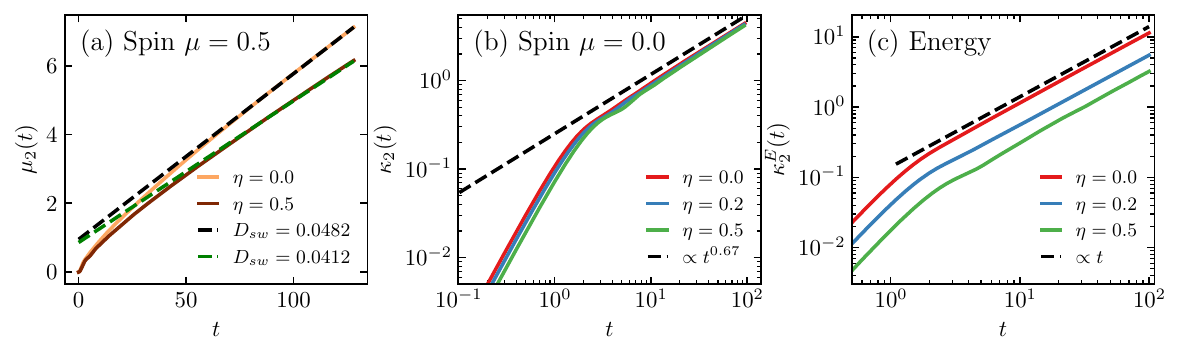}
    \caption{\textbf{Second moments and cumulants of the magnetization and energy transfer.} Panel (a) shows the second moment of the net magnetization transfer through the middle of the system, $\mu_2(t)=\langle\left[S_{\rm left}^z(t)-S_{\rm left}^z(0)\right]^2\rangle$. The orange and dark red solid lines correspond to $\mu_2(t)$ computed at finite chemical potential $\mu=0.5$, for $\eta=0$ and $\eta=0.5$ respectively. The black and green dashed lines correspond to $\sim D_{\rm sw}t$, with the hydrodynamic predictions for the Drude self-weights $D_{\rm sw}$ reported in Section~\ref{sec:drude-ballistic}. Panel (b) shows the second cumulant of the net magnetization transfer through the middle of the system, $\kappa_2(t)$, for different values of the model's parameter $\eta$, at zero chemical potential ($\mu=0$). We observe the asymptotic scaling $\kappa_2(t)\sim t^{2/3}$, implying the dynamical exponent $z=3/2$. Panel (c) shows the second cumulant $\kappa^{E}_2(t)$ of the net energy transfer through the middle of the system, at infinite temperature and for different values of $\eta$. Irrespectively of the model's parameter $\eta$, we find $\kappa^{E}_2(t) \sim t$, indicating ballistic energy transport.}  
    \label{fig:spin_transport}
\end{figure*}

\subsection{Quantum generating function}

The charge moments are obtained using the quantum generating function (QGF) method, following Ref.
~\cite{valli2024efficientcomputationcumulantevolution}.
We consider observables of the form $\mc{O}_\mc{S} = \sum_{j \in \mc{S}} \mc{O}_j$ that correspond to the total charge in region $\mc{S}$. Within a tensor-network framework, we start by constructing the matrix-product operator (MPO) representation of
\begin{align}
    R(\lambda,0)=\ee^{\ii \lambda \mc{O}_\mc{S}(0)},
\end{align}
with a complex generating parameter, $\lambda \in \mathbb{C}$. We point out that for single-site observables, such as the spin magnetization $\mc{O}_j=S_j^z$, the MPO representation of $R(\lambda,0) = \bigotimes_{j \in \mc{S}} \ee^{\ii \lambda \mc{O}_j}$ has bond dimension $\chi = 1$. Energy density operators, however, have larger support and cannot generally be written as tensor products of single-site operators. 
The time evolution is then given by
\begin{align}
    R(\lambda,t)
    =U^\dagger(t) R(\lambda,0) U(t),
\end{align}
and it is implemented using a time-evolving block decimation (TEBD) architecture. 
The algorithm efficiency is largely controlled by the size of $\lambda$: for instance,  for small enough $|\lambda|$, the operator $R(\lambda,t)=\ee^{\ii \lambda \mc{O}_\mc{S}(t)}$ may remain sufficiently close to the identity at all times.
An important and subtle bottleneck is  machine precision, as the extraction of higher-order moments is directly tied to high powers of $|\lambda|=\lambda_0$. Unless stated otherwise, we use $\lambda_0=0.03$ and a TEBD time step $\dd t=0.1$ in the numerical simulations.

For a grand-canonical state $\varrho_\mu$, Eq.~\eqref{eq:grand-canonical-state}, the moment-generating function is then obtained as
\begin{align}
\label{eq:moment_generating function}
    G_\mathrm{QGF}(\lambda,t) = \mathrm{Tr}[R(\lambda,t) R^\dagger(\lambda^*,0) \varrho_\mu].
\end{align}
The charge moments are extracted by taking discrete derivatives of $G_\mathrm{QGF}(\lambda,t)$. For this, we consider $\lambda=\lambda_0\ee^{\ii\phi} $ and run several simulations for different phases $\phi\in\{0,\pi/4,-\pi/4,\pi/2\}$ of the generating parameter, and combine them as follows \cite{valli2024efficientcomputationcumulantevolution}
\begin{align}
    \mu_{2}(t)+O(\lambda_0^8)
    &=-\frac{2!}{4\lambda_0^2}
    \Big[\Im G_{\frac{\pi}{4}}(\lambda_0,t)-\Im G_{-\frac{\pi}{4}}(\lambda_0,t)\nonumber\\
    &-\Re G_{\frac{\pi}{2}}(\lambda_0,t)+\Re G_{0}(\lambda_0,t)\Big],\nonumber\\
    \mu_{4}(t)+O(\lambda_0^8)
    &=-\frac{4!}{4\lambda_0^4}
    \Big[\Re G_{\frac{\pi}{4}}(\lambda_0,t)+\Re G_{-\frac{\pi}{4}}(\lambda_0,t)\nonumber\\
    &-\Re G_{\frac{\pi}{2}}(\lambda_0,t)-\Re G_{0}(\lambda_0,t)\Big],
\end{align}
where we adopt the shorthand notation  $G_{\phi}(\lambda_0,t)$ to highlight phase dependence.
Finally, we mention that in computing the generating function $G_\mathrm{QGF}(\lambda,t)$, one performs the direct contraction between $R(\lambda,t)$ and $R^\dagger(\lambda^*,0)$, which differs from the operator expression $\ee^{\ii\lambda[\mc{O}_\mc{S}(t)-\mc{O}_\mc{S}(0)]}$ since $\mc{O}_\mc{S}(t)$ and $\mc{O}_\mc{S}(0)$ do not generally commute. The resulting discrepancy, defined as $\varepsilon(t) = |G_\mathrm{QGF}(\lambda,t) - G(\lambda,t)|$, can be expanded in powers of $\lambda$. The leading nonvanishing contributions are of order $O(\lambda^4)$ and therefore do not influence the computation of the second (or third) moments, although they may affect the estimation of the fourth moment $\mu_4(t)$. 
Nevertheless, for the simulations presented here, the difference $\varepsilon(t)$ introduces only subleading time-dependent corrections, which are washed out in the long-time estimates of the excess kurtosis, defined as $\gamma_4(t)=\mu_{4}(t)/\mu_{2}^{2}(t)-3$.

\subsection{Fluctuations of magnetization and energy}
\label{sec:spin}

To validate our approach, we first perform numerical simulations at a nonzero chemical potential. Spin transport is probed through magnetization in the left-half of the system, defined as $\mc{O}_\mc{S}=S_\mrm{left}^{z}=\sum_{j=1}^{L/2}S_{j}^{z}$. We simulate a finite chain of length $L=256$ at a chemical potential $\mu=0.5$. The maximum bond dimension during the time evolution is $\chi=512$. The time dependence of the second spin-transfer cumulant $\kappa_{2}(t)$ is shown in Fig.~\hyperref[fig:spin_transport]{2(a)} for two coupling values, $\eta=0$ and $\eta=0.5$. The numerical results are contrasted with the long-time hydrodynamic predictions reported in Section~\ref{sec:drude-ballistic} (dashed lines).

We now turn to the half-filling limit ($\mu\to0$), where superdiffusion is expected to emerge, according to the predictions of Section~\ref{sec:superdiffusion-self-consistent}.
To inspect this prediction, we consider the second cumulant, $\kappa_2(t)$. Note that it coincides with the second moment, since the first moment vanishes identically. Numerical simulations are performed for parameters $\eta=0$, $0.2$, and $0.5$, for a system of size $L=256$. Within the tensor-network algorithm we set the maximum bond dimension to $\chi=512$.
The extracted second moments are presented in Fig.~\hyperref[fig:spin_transport]{2(b)}. We find that spin transport is consistent with superdiffusion with dynamical exponent $z=3/2$, irrespective of the value of $\eta$, i.e., 
irrespective of how strongly $J_\chi$ enters the dynamics.
As $J_\chi$ parametrizes the strength of the time-reversal breaking term, this suggests that violation of time-reversal does not alter the universal properties the superdiffusive regime, as far as second-order correlations are concerned. 

Before considering higher-order spin moments, we further consider the second cumulant of the energy transfer, $\kappa_{2}^E(t)$.
To capture the transfer of energy we set $\mathcal{O}_\mc{S}=H_{\rm left}$, where is the truncation of the Hamiltonian~\eqref{eq:hamiltonian} to the left half of the ladder.
In Fig.~\hyperref[fig:spin_transport]{2(c)} we show $\kappa_{2}^{E}(t)$, for system size $L=128$ and various values of the model's parameter $\eta$. Irrespective of the value of $\eta$, we observe linear growth of the second cumulant $\kappa_{2}^{E}(t)\propto t^{1/z_E}$, corresponding dynamical exponent being $z_E=1$ characteristic of ballistic transport.
Finally, we confirm non-Abelian symmetries alone are not enough to support superdiffusion by moving away from the integrable line. We consistently observe (not shown) diffusion of spin and energy when considering the dynamics of $H_\chi$ alone.

We next focus on how the breaking of time-reversal symmetry affects the higher moments, restricting the discussion to magnetization transfer. Firstly, we note that, due to the $C_2$ symmetry, i.e., the link parity $\mathcal{P}_\ell$ rotating the ladder around the central link, the odd moments corresponding to the magnetization transfer vanish in the grand-canonical state, Eq.~\eqref{eq:grand-canonical-state}. Specifically, we have:
\begin{equation}\mu_n(t)=\langle\left[S^z_{\rm left}(t)-S^z_{\rm left}(0)\right]^n\rangle=0
\end{equation}
for odd $n$. This follows from the fact that, on the one hand we have $S^z_{\rm left}(t)-S^z_{\rm left}(0)=-S^z_{\rm right}(t)+S^z_{\rm right}(0)$, due to the conservation of the total spin $S^{z}_{\rm total}=S^z_{\rm left}+S^z_{\rm right}$, and on the other hand  $\mathcal{P}_{\ell}\left[S^z_{\rm left}(t)-S^z_{\rm left}(0)\right]=S^z_{\rm right}(t)-S^z_{\rm right}(0)$, as well as $\langle\mathcal{P}_{\ell}[\bullet]\rangle=\langle\bullet\rangle$ due to the $C_2$ invariance of the state. Expressing odd cumulants of the transferred magnetization in terms of moments, one finds that each term contains an odd moment, and the odd cumulants therefore vanish as well. Since, by the continuity equation, the net magnetization transferred from the left- to the right-half of the system is equal to the time-integrated current flowing through the middle of the system, we conclude the full counting statistics of the spin current obeys fluctuation symmetry.

\begin{figure}[t]
    \centering
    \includegraphics[width=1.\linewidth]{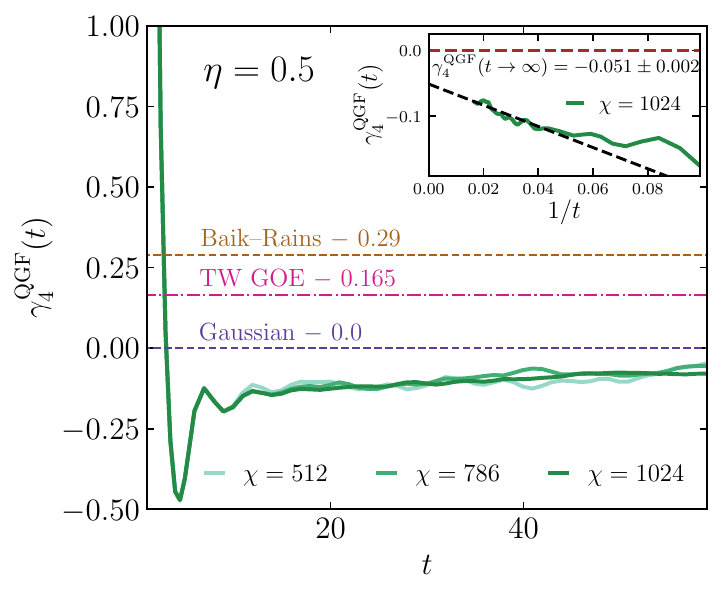}
    \caption{\textbf{Higher-order fluctuations corresponding to the magnetization transfer.} Plotted is the excess kurtosis $\gamma_4^\mathrm{QGF}(t)$ of the magnetization transfer for the chiral ladder with $L=128$ sites and $\eta=1/2$. Various bond dimensions $\chi$ are used. The dashed lines correspond to the predictions within the KPZ universality class (either Baik-Rains or Tracy-Widom), and to Gaussian fluctuations. The inset shows $\gamma_4^\mathrm{QGF}(t)$ as a function of $1/t$ for $\chi=1024$. Extrapolating to $t\to \infty$ gives a negative and non-zero value of the excess kurtosis, suggesting a non-Gaussian nature of the fluctuations.}
    \label{fig:excess_kurtosis}
\end{figure}

We now analyse the long-time behaviour of the excess kurtosis, $\gamma_4(t)$, which quantifies deviations from Gaussian fluctuations, for which $\gamma_4(t)=0$. In the KPZ universality class, the excess kurtosis is expected to be positive with typical values approaching $\gamma_4\approx0.29$ and $\gamma_4\approx0.165$~\cite{PrahoferSpohn}, corresponding respectively to the Baik–Rains~\cite{baik2000limiting} and 
Tracy–Widom Gaussian orthogonal ensemble (TW-GOE) distributions~\cite{Tracy1994}, depending on the initial conditions on the height function. These reference values are indicated by the dashed horizontal lines in Fig.~\ref{fig:excess_kurtosis}.
To compute $\gamma_4^\mathrm{QGF}(t)$, we simulate a system of length $L=128$ at coupling $\eta=0.5$, using several maximum bond dimensions $\chi=512$, $786$, and $1024$. The resulting data are shown in Fig.~\ref{fig:excess_kurtosis}. Despite the second moment exhibiting a KPZ-type dynamical exponent, the excess kurtosis remains inconsistent with KPZ predictions: $\gamma_4^\mathrm{QGF}(t)$ attains small negative values at long times. The discrepancies visible in Fig.~\ref{fig:excess_kurtosis} are amplified by error propagation in the evaluation of $\gamma_4^\mathrm{QGF}(t)$, which involves dividing $\mu_4(t)$ by $\mu_2^2(t)$. Both $\mu_2(t)$ and $\mu_4(t)$ are, however, well converged across different bond dimensions, as shown in the Appendix~\ref{app:convergence}.
The inset of Fig.~\ref{fig:excess_kurtosis} displays a simple $1/t$ extrapolation toward $t\to\infty$, yielding a small negative asymptotic value, $\gamma_4^\mathrm{QGF}(t\to\infty)\approx-0.051\pm0.002$. While the precise numerical value, as well as the reported error bar (that, here, just includes error propagation from the linear fit), should be taken {\it cum grano salis} and possibly sharpened by future large scale simulations, these results strongly indicate that the observed fluctuations deviate from KPZ behavior, similarly with recent findings in the XXX model~\cite{valli2024efficientcomputationcumulantevolution,rosenberg2024dynamics}.

\section{Chiral unitary circuit}
\label{sec:circuit}

We now introduce an integrable brickwork-type unitary circuit, designed to preserve the key symmetries of the continuous-time evolution generated by the Hamiltonian in Eq.~\eqref{eq:hamiltonian}. The key question that we want to address here is, whether giving up energy conservation can dramatically affect the overall picture obtained in the case of Hamiltonian dynamics, i.e., whether a non-KPZ kurtosis and $z>1$ still appear in the absence of time-reversal symmetry. Furthermore, this circuit formulation of the dynamics is naturally suited for implementation in digital simulators, as demonstrated in Refs.~\cite{Kalinowski2023NonAbelian,Evered2025Probing}, where chiral three-body terms are engineered. 

The Hamiltonian in Eq.~\eqref{eq:hamiltonian} can be rewritten as $H=\sum_{j=1}^{L/2}h_{[2j-1,2j+2]}$, where $[a,b]$ denotes the (closed) interval of integers between $a$ and $b$, and
\begin{align}
    h_{[1,4]}\!=\,&J_1 \mathbf{S}_{1}\!\cdot\!\mathbf{S}_{2}\!+\!J_1 \mathbf{S}_{2}\!\cdot\!\mathbf{S}_{3}\!+\!J_2 \mathbf{S}_{1}\!\cdot\!\mathbf{S}_{3}\!+\!J_2\mathbf{S}_{2}\!\cdot\! \mathbf{S}_{4}\notag\\
    &+\!J_\chi \mathbf{S}_{1}\!\cdot\!(\mathbf{S}_{2}\!\cross\! \mathbf{S}_{3})\!-
    \!J_\chi \mathbf{S}_{2}\!\cdot\!(\mathbf{S}_{3}\!\cross \!\mathbf{S}_{4}),
    \label{eq:H-density}
\end{align}
is the four-site local Hamiltonian density, parametrized by $J_{1,2}$ and $J_\chi$ given in Eq.~\eqref{eq:integrable_line}. Time-discretization is based on the Trotter-Suzuki formula
\begin{align}
    \ee^{-\ii t H}\!=\!\lim_{n\to\infty}\!\Big(\!\ee^{-\ii\frac{t}{n}\!H_{\rm A}}\ee^{-\ii\frac{t}{n}\!H_{\rm B}}\!\Big)^n,
    \label{eq:trotter-suzuki}
\end{align}
with $H_{\rm A}\!=\!\sum_{j=1}^{L/4}h_{[4j-1,4j+2]}$ and $H_{\rm B}\!=\!\sum_{j=1}^{L/4}h_{[4j-3,4j]}$. The right-hand side of Eq.~\eqref{eq:trotter-suzuki} factorizes into a sequence of four-site unitary gates, each one corresponding to the exponential of the local density in Eq.~\eqref{eq:H-density}. To preserve integrability, we would moreover like the unitary gate to satisfy the Yang-Baxter equation~\cite{vanicat2018,paletta2025}. For the chiral spin ladder, this is only possible by choosing the unitary gate $U_{[1,4]}$ which matches $\exp\{-(\ii t/n) h_{[1,4]}\}$ up to an $O([t/n]^3)$ correction. In particular, we have to set
\begin{align}
    U_{[1,4]}\!=\!\check{R}_{2,3}(\tau\!-\!\delta)\check{R}_{1,2}(\tau)\check{R}_{3,4}(\tau)\check{R}_{2,3}(\tau\!+\!\delta),
    \label{eq:unitary-gate}
\end{align}
where $\check{R}_{1,2}(\lambda)=(\mathds{1}+\ii\lambda P_{1,2})/(1+\ii\lambda)$
is the Yang-Baxter gate in the integrable XXX circuit~\cite{vanicat2018}, $P_{1,2}=\tfrac{1}{2}\mathds{1}+2\mathbf{S}_1\!\cdot\mathbf{S}_2$ denoting the permutation of two spins $1/2$, and $\delta\!=\!-\!\sqrt{\eta/(1\!-\!\eta)}$. For the unitary gate in Eq.~\eqref{eq:unitary-gate}, we find 
$U_{[1,4]}\!-\!\exp\{\ii \tau [4h_{[1,4]}\!+\!(2\!-\!\eta)\mathds{1}]\}\!=\!O(\tau^3)$ (see Appendix~\ref{app:circuit-gate}). Setting $\tau=-t/n$, we can then reproduce the continuous-time dynamics of the chiral ladder in the Trotter limit $\ee^{-\ii t H}=\lim_{n\to\infty}\mathds{U}^{n}$,
where
\begin{align}
    \mathbb{U}
    =\prod_{j=1}^{\frac{L}{4}}U_{[4j-1,4j+2]}
    \prod_{j=1}^{\frac{L}{4}}U_{[4j-3,4j]}
    \label{eq:brickwork-circuit}
\end{align}
is the one-step propagator. Its circuit diagram is depicted in Fig.~\hyperref[fig:ladder]{1(b)}.

\begin{figure}[t]
    \centering
    \includegraphics[width=1.\linewidth]{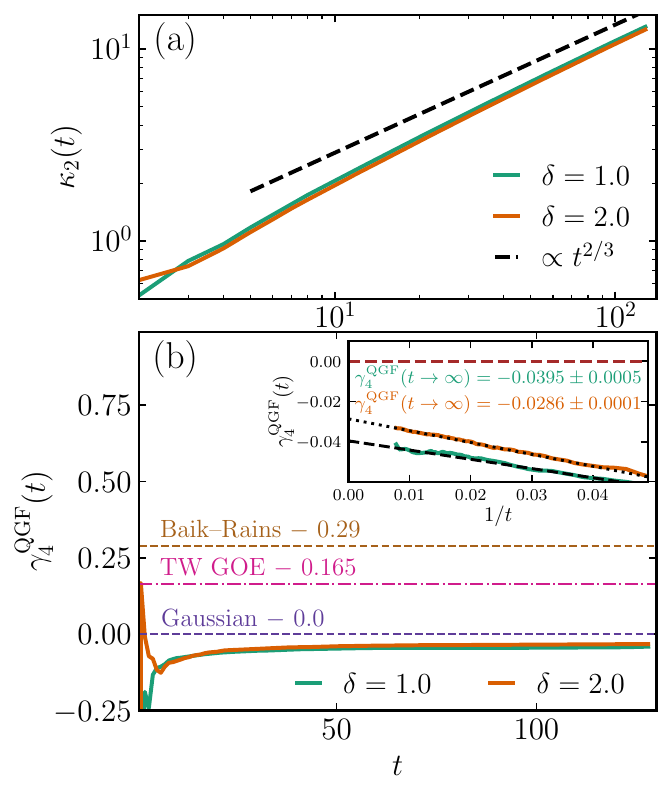}
    \caption{\textbf{Spin transport and higher-order fluctuations in the chiral circuit.}
    Panel (a) shows the second moment $\mu_{2}(t)$ for $\delta=1$ and $2$. The dashed line indicating the power-law scaling $\propto t^{2/3}$ serves as a guide to the eye.
    Panel (b) shows the excess kurtosis $\gamma_4^\mathrm{(QGF)}$ for the same values of $\delta$. Dashed lines show theoretical predictions for the standard KPZ (TW and BR) distributions, as well as for the Gaussian fluctuations. The inset highlights the $1/t$ infinite-time extrapolation. Results were computed using the QGF method with maximum bond dimension $\chi=1024$, for a system of size $L=256$, and circuit step $\tau=1$.}
    \label{fig:circuit-moments}
\end{figure}

To characterize spin transport in the circuit, we apply the QGF method implementing the one-step propagator $\mbb{U}$ in Eq.~\eqref{eq:brickwork-circuit} with TEBD
~\footnote{We implement open boundary conditions by removing one of the four-site unitary gate, such that $\mathbb{U}_\mrm{OBC}=\prod_{j=1}^{\frac{L}{4}-1}U_{[4j-1,4j+2]}\prod_{j=1}^{\frac{L}{4}}U_{[4j-3,4j]}$}.
Throughout, we set $\tau=1$ and examine two values of the coupling parameter, $\delta=1$ and $\delta=2$. The simulations are performed for a system of size $L=256$ and a maximum bond dimension $\chi=1024$. The second cumulant $\kappa_{2}(t)$ is shown in Fig.~\hyperref[fig:circuit-moments]{4(a)} and exhibits a power-law growth consistent with the superdiffusive exponent $z=3/2$. 
However, the corresponding excess kurtosis $\gamma_4^\mathrm{QGF}(t)$, displayed in Fig.~\hyperref[fig:circuit-moments]{4(b)}, remains negative at all accessible late times. 
The inset, plotted versus $1/t$, indicates only a slow drift toward larger values, with an infinite-time extrapolation yielding $\gamma_4^\mathrm{(QGF)}(t\to\infty,\delta=1)\approx-0.0395\pm 0.0005$ and $\gamma_4^\mathrm{(QGF)}(t\to\infty,\delta=2)\approx-0.0286\pm 0.0001$. 
The circuit results are thus compatible with the ones obtained in the continuous time evolution, presenting slightly negative excess kurtosis values, reinforcing the non-KPZ nature of fluctuations observed so far.

\section{Conclusions}
\label{sec:conclusions}

In this work we have considered several aspects of transport in an integrable SU(2)-invariant spin ladder with tunable chiral terms breaking the time-reversal symmetry. We have found that, even in the absence of time reversal, the spin transport at infinite temperature and zero chemical potential is superdifussive, with dynamical critical exponent $z=3/2$, but with fluctuations of the net magnetization transfer that are not consistent with the Kardar-Parisi-Zhang universality class. Surprisingly, we also find that the statistics of the associated net spin current obeys fluctuation symmetry despite the lack of time reversal. Time reversal symmetry is therefore sufficient~\cite{doyon-perfetto-sasamoto-yoshimura-2023}, but not a necessary condition for the fluctuation symmetry to appear. In the case of the chiral spin ladder, the latter instead appears due to a particular geometrical symmetry of the ladder.

Firstly, our findings establish that the anomalous superdiffusion can be robust to the time-reversal symmetry breaking. Furthermore, they suggest that the mere absence of time-reversal symmetry is insufficient for the stabilization of KPZ-type superdiffusion, even though the KPZ universality class itself also lacks that symmetry~\cite{NonlinearFH_JDN}. Finally, we have introduced an integrable chiral quantum circuit exhibiting the same symmetries as the chiral ladder, apart from energy conservation. Here, the results regarding the spin transport were consistent with the ones in the continuous-time model. The chiral spin circuit introduced herein provides a natural platform for experimental investigation of the effects of broken space-time symmetries on the transport phenomena~\cite{Kalinowski2023NonAbelian,Evered2025Probing}.

\section*{Acknowledgments} 
We thank B.\,K. Agarwalla, J. Goold, E. Ilievski, A. Nersesyan, and A. Scardicchio for discussions. 
M.\,D. was partly supported by the QUANTERA DYNAMITE PCI2022-132919, by the EU-Flagship programme Pasquans2, by the PNRR MUR project PE0000023-NQSTI, the PRIN programme (project CoQuS), and the ERC Consolidator grant WaveNets (Grant agreement ID: 101087692). L.\,Z. acknowledges support by the Slovenian Research Agency (ARIS) under Research Startup Program No. SN-ZRD/22-27/0510—NESY and Research Program P1-0402, as well as the European Research Council (ERC) under Advanced Grant No. 101096208—QUEST. D.\,S.\,B. acknowledges the CINECA Grant under the ISCRA-C (HP10C66VX2) program. H.\,B.\,X. was supported by the MIUR programme FARE (MEPH).
Simulations were performed using the Julia versions of ITensor~\cite{Matthew2022itensor,Matthew2022codebase} and QuSpin~\cite{philip2017quspin,philipe2019quspin} libraries.

\appendix

\section{Integrability, thermodynamics, and integrable chiral circuit}
\label{app:integrability}

Here, we discuss integrability of the chiral spin ladder Hamiltonian~\eqref{eq:hamiltonian}. In Appendix~\ref{app:transfer-matrices} we introduce the transfer matrix used in the algebraic Bethe ansatz to diagonalize the Hamiltonian~\cite{Frahm1996Integrable}. We also relate the Hamiltonian to the conserved charges of the (integrable) Trotterized Heisenberg model~\cite{vanicat2018}. This correspondence allows us to use the existing Bethe ansatz results for integrable circuits (see, e.g., Ref.~\cite{paletta2025,Ratchets}) in the description of the chiral spin ladder's thermodynamics. The latter is summarized in Appendix~\ref{app:TBA}: it is the essential ingredient in the hydrodynamic description of the transport. Lastly, in Appendix~\ref{app:circuit-gate} we describe how to obtain the unitary gate that can be used in the integrable Trotterization of the chiral spin ladder.

\subsection{Transfer matrix of the chiral spin ladder}
\label{app:transfer-matrices}

Integrability of the chiral spin ladder Hamiltonian~\eqref{eq:hamiltonian} can be traced back to the SU(2) symmetric $R$ matrix of the XXX model,
\begin{align}
    R_{1,2}(\lambda)=\frac{\lambda\mathds{1}-\ii P_{1,2}}{\lambda-\ii}.
    \label{eq:R-matrix}
\end{align}
Here, $P_{1,2}=\frac{1}{2}\mathds{1}+2\mathbf{S}_1\!\cdot\mathbf{S}_2$ is a permutation acting on two spins $1/2$, and $\lambda\in\mathds{C}$ is a free spectral parameter. For any two $\lambda,\vartheta\in\mathds{C}$, the $R$ matrix satisfies the Yang-Baxter equation
\begin{align}
    R_{1,2}(\lambda-\vartheta)&R_{1,3}(\lambda)R_{2,3}(\vartheta)=\notag\\
    &=R_{2,3}(\vartheta)R_{1,3}(\lambda)R_{1,2}(\lambda-\vartheta).
    \label{eq:yang-baxter}
\end{align}

We use the $R$ matrix to define a family of transfer matrices,
\begin{align}
    \label{eq:transfer-matrix}
    T_\kp(\lambda)={\rm Tr}_{a}\!\left[\prod_{1\le x\le L/2}^{\rightarrow} R_{a,2x-1}(\lambda)R_{a,2x}(\lambda\!-\!\kp)\right],
\end{align}
where ${\rm Tr}_a(\bullet)$ is a partial trace over the auxiliary spin $1/2$. The arrow above the product indicates the ordering of the physical spins labelled by $1,2,\ldots,L$. The Yang-Baxter equation~\eqref{eq:yang-baxter} implies
\begin{align}
    [T_\kp(\lambda),T_\kp(\vartheta)]=0
\end{align}
for a fixed $\kp\in\mathds{R}$, and for any pair of spectral parameters $\lambda,\vartheta\in\mathds{C}$. The transfer matrices can be simultaneously diagonalized by means of the algebraic Bethe ansatz~\cite{faddeev1996}.

The integrable chiral spin ladder Hamiltonian~\eqref{eq:hamiltonian}, whose parameters $(J_1(\eta),J_2(\eta),J_\chi(\eta))$ lie on  a curve specified in Eq.~\eqref{eq:integrable_line}, is  obtained as~\cite{Frahm1996Integrable}
\begin{align}
    H=\frac{1}{4\ii}\frac{\rm d}{{\rm d}\lambda}\log\left[T_\kp(\lambda+\kp)T_\kp(\lambda)\right]\bigr|_{\lambda=0},
    \label{eq:H-from-T}
\end{align}
provided that we set $\kp=-\sqrt{\eta/(1-\eta)}$. Besides the Hamiltonian, which generates translations in time, the two-site lattice translations can likewise be expressed using the transfer matrix~\eqref{eq:transfer-matrix} as
\begin{align}
    \label{eq:translations}
    T_\kp(\kp)T_\kp(0)=(P_{1,2}P_{1,3}\cdots P_{1,L})^2\equiv \ee^{-\ii 2 \mathcal{P}}.
\end{align}
We use this equation as a definition of the momentum operator, $\mathcal{P}$, which generates two-site translations in space (we normalize $\mathcal{P}$ with a prefactor $2$, reflecting the two-site translation). Bethe ansatz gives access to the eigenvalues of $T_\kp(\lambda)$, and thus to the energies and (quasi)momenta.

We note that the staggered transfer matrix~\eqref{eq:transfer-matrix} also plays the central role in the construction of integrable Trotterization of the XXX model~\cite{vanicat2018}. The latter is an integrable quantum unitary circuit that, in the so-called Trotter-Suzuki limit, reproduces the continuous-time evolution with the isotropic spin-$1/2$ Heisenberg Hamiltonian. In fact, the chiral spin ladder Hamiltonian~\eqref{eq:H-from-T} can be expressed as
\begin{align}
    H=\frac{1}{4\ii}(Q^+_1+Q^-_1),
\end{align}
where 
\begin{align}
    Q_1^{\pm}\equiv \frac{\rm d}{{\rm d}\lambda}\log T(\lambda+\tfrac{\kp}{2};\kp)\bigr|_{\lambda=\pm\tfrac{\kp}{2}}
    \label{eq:first-two-charges}
\end{align}
are precisely the first two local conserved charges of the integrable XXX circuit---cf. Eqs.~(9), (10), and~(11) in Ref.~\cite{vanicat2018}. It follows that the chiral spin ladder Hamiltonian and the integrable Trotterization of the Heisenberg spin chain share the same (quasi)local conserved charges. This means that, following a quantum quench, the chiral spin ladder Hamiltonian and the integrable XXX circuit evolve the initial state towards a steady state whose local properties are captured by the same generalized Gibbs ensemble (GGE).

\subsection{Quasiparticle content of a thermodynamic state}
\label{app:TBA}

The eigenvalues of local charges, such as the Hamiltonian in Eq.~\eqref{eq:H-from-T}, are extensive. They are sums of contributions from quasiparticles which correspond to bound states of $s$ magnons ($s\in\mathds{N}$), each magnon carrying spin $1$ (a magnon is associated with a spin flip). Extensivity allows us to express the density of a local charge $Q$ in the thermodynamic state as
\begin{align}
    \lim_{N,L\to\infty\atop N/L\text{ finite}}\frac{\langle Q\rangle}{L}=\sum_{s=1}^\infty\int{\rm d}\lambda \,\rho_s(\lambda)q_s(\lambda).
    \label{eq:charge-density}
\end{align}
Here, $\rho_s(\lambda)$ is the density of quasiparticles, and $q_s(\lambda)$ is the bare charge carried by the type-$s$ quasiparticle with rapidity $\lambda$. For example, denoting the rapidity derivative by $(\bullet)'$, the bare energy reads
\begin{align}
    \label{eq:energy}
     \varepsilon_s(\lambda)=-\frac{1}{2}p_s'(\lambda).
\end{align}
Here, due to the staggered structure of the transfer matrix~\eqref{eq:transfer-matrix}, the bare momentum $p_s(\lambda)$ can be expressed as a sum of bare momenta $\tilde{p}_s(\lambda)$ and $\tilde{p}_s(\lambda-\kp)$ in the XXX model. Specifically, we have
\begin{align}
    p_s(\lambda)=\tilde{p}_s(\lambda)+\tilde{p}_s(\lambda-\kp),
\label{eq:momentum}
\end{align}
with
\begin{align}
    \tilde{p}_s(\lambda)\equiv\frac{\ii}{2}\log\left(\frac{\lambda+\ii \frac{s}{2}}{\lambda-\ii \frac{s}{2}}\right).
\end{align}
The same decomposition of the momentum occurs in the integrable Trotterization of the XXX model---see Ref.~\cite{paletta2025}.

\subsubsection{Quasiparticle density}

The quasiparticle density entering Eq.~\eqref{eq:charge-density} is customarily factorized as $\rho_s(\lambda)=n_s(\lambda)\rho^{\rm t}_s(\lambda)$,
where $n_s(\lambda)$ is the occupancy ratio and $\rho^{\rm t}_s(\lambda)$ is the total state density, i.e., the density of particles and holes. The equations that determine $n_s(\lambda)$ and $\rho_s^{\rm t}(\lambda)$ are the thermodynamic Bethe-ansatz (TBA) equations. They appear in the thermodynamic limit of the equations which characterize the eigenvalues of the transfer matrix~\eqref{eq:transfer-matrix}. 

In this work we are mostly interested in the thermodynamic state at an infinite temperature $T$ and a finite chemical potential $\mu$, Eq.~\eqref{eq:grand-canonical-state}, for which the TBA equations read~\cite{paletta2025,takahashi1999}
\begin{align}
    &\log(n_s^{-1}\!-\!1)\!=\!\mu s\!-\!\!\sum_{\ell=0}^\infty \!\int\!\!{\rm d}\vartheta  K_{s,\ell}(\lambda\!-\!\vartheta)\!\log(1\!-\!n_\ell),
    \label{eq:TBA-aligns-1}
    \\
	&\rho^{\rm t}_s(\lambda)+\sum_{\ell=0}^\infty\int{\rm d}\vartheta K^{}_{s,\ell}(\lambda\!-\!\vartheta)n^{}_{\ell}\rho_{\ell}^{\rm t}(\vartheta)=\frac{p_s'(\lambda)}{2\pi}.
    \label{eq:TBA-aligns-2}
\end{align}
Here, $K_{s,\ell}(\lambda-\vartheta)$ is the scattering kernel with an explicit form~\cite{takahashi1999}
\begin{align}
    K_{s,m}(\lambda)\!=\!\frac{1}{\pi}\!\sum_{j=\frac{|s-m|}{2}}^{\frac{s+m}{2}-1}\!\left(\frac{j}{\lambda^2\!+\!j^2}\!+\!\frac{j\!+\!1}{\lambda^2\!+\!(j\!+\!1)^2}\right).
	\label{eq:kernel}
\end{align}

From Eq.~\eqref{eq:TBA-aligns-1} one obtains a rapidity-independent occupancy ratio $n_s$, reported in Eq.~\eqref{eq:occupancy-ratio}---see, e.g., Ref.~\cite{takahashi1999}. Since $n_s$ does not depend on the rapidity, Eq.~\eqref{eq:TBA-aligns-2} is a linear equation for the total state density $\rho^{\rm t}_s(\lambda)$. As a consequence, since the momentum~\eqref{eq:momentum} decomposes into two parts, so does $\rho_s^{\rm t}(\lambda)$. Similarly as in integrable circuits~\cite{paletta2025}, we then find 
\begin{align}
    \rho_s^{\rm t}(\lambda)=\tilde{\rho}^{\rm t}_s(\lambda)+\tilde{\rho}^{\rm t}_s(\lambda-\kp),
\end{align}
where $\tilde{\rho}^{\rm t}_s(\lambda)$, reported in Eq.~\eqref{eq:state-density}, satisfies the TBA equation~\eqref{eq:TBA-aligns-2} in the XXX model~\cite{ilievski2018}.

\subsubsection{Dressed charges and effective velocities}

Equation~\eqref{eq:TBA-aligns-2} is a particular example of the dressing equation~\cite{doyon2020lecture},
\begin{align}
    q_s^{\rm dr}(\lambda)+\sum_{\ell=0}^\infty\int{\rm d}\vartheta K^{}_{s,\ell}(\lambda-\vartheta)n^{}_{\ell}q_\ell^{\rm dr}(\vartheta)=q_s(\lambda),
    \label{eq:dressing}
\end{align}
which determines how the bare charge carried by the quasiparticles changes in the presence of interactions: $q_s\mapsto q_s^{\rm dr}$. In particular, according to Eq.~\eqref{eq:TBA-aligns-2} we have $(p_s')^{\rm dr}(\lambda)=2\pi\rho^{\rm t}_s(\lambda)$. Below, we summarize also the derivation of the dressed energy derivative $(\varepsilon_s')^{\rm dr}$, which enters the effective velocity $v_s=(\varepsilon_s')^{\rm dr}/(p_s')^{\rm dr}$ alongside $(p_s')^{\rm dr}$, and the dressed charge---magnetization. We will stick with notation $q_s^{\rm dr}$ for the latter.

First, we obtain the dressing equation for the magnetization, which encodes the mapping $s\mapsto q_s^{\rm dr}$, by applying $\partial_\mu$ to Eq.~\eqref{eq:TBA-aligns-1}. Comparing the result to Eq.~\eqref{eq:dressing}, we then identify the dressed magnetization reported in Eq.~\eqref{eq:charge}.

Second, applying $\partial_\lambda$ to Eq.~\eqref{eq:TBA-aligns-2}, and noting that $\lim_{\lambda\to\pm\infty}K_{s,m}(\lambda)=0$, we obtain 
\begin{align}
    \partial^{}_\lambda\rho_s^{\rm t}(\lambda)+\sum_{m}\!\!\int\!\!{\rm d}\vartheta K^{}_{s,m}(\lambda\!-\!\vartheta)n^{}_{m}\partial^{}_\vartheta\rho_s^{\rm t}(\vartheta)\!=\!\frac{p_{s}''(\lambda)}{2\pi}.
\end{align}
Recalling Eq.~\eqref{eq:energy}, we now recognize
\begin{align}
    (\varepsilon_s')^{\rm dr}(\lambda)=-\pi \partial_\lambda\rho_s^{\rm t}(\lambda).
\end{align}
The dressed momentum and energy derivatives enter the hydrodynamic mode velocity
\begin{align}
    v_s(\lambda)=\frac{(\varepsilon_s')^{\rm dr}(\lambda)}{(p_s')^{\rm dr}(\lambda)},
\end{align}
whose final form is reported in Eq.~\eqref{eq:velocity}.

\subsection{Integrable circuit for the chiral spin ladder}
\label{app:circuit-gate}

In the context of the integrable XXX circuit~\cite{vanicat2018}, the local Hamiltonian density $h_{[1,4]}$ from Eq.~\eqref{eq:H-density} enters the boost operator $B$, which generates the hierarchy of conserved charges. Specifically, 
\begin{align}
B=\sum_j j  h_{[2j-1,2j+2]},
\end{align}
which is defined in a thermodynamically-large system, translates within the hierarchy of charges of the integrable circuit according to the boost relation $[B,Q_n^\pm]= Q_{n+1}^\pm$. Here, the first two charges among $Q^{\pm}_n$ are $Q_1^\pm$ from Eq.~\eqref{eq:first-two-charges}. 

In Ref.~\cite{vanicat2018} it was shown that one can obtain the local density $h_{[1,4]}$, entering the boost operator $B$, as
\begin{align}
   h_{[1,4]}=\frac{1}{4\ii}\partial_\lambda\mathds{R}_{[1,4]}(\lambda;\delta)\bigr|_{\lambda=0}-\frac{2-\eta}{4}\mathds{1},
\end{align}
where $\delta=-\sqrt{\eta/(1-\eta)}$ and
\begin{align}
    \mathds{R}_{[1,4]}(\lambda;\delta)\!=\!\check{R}_{2,3}(\lambda\!-\!\delta)\check{R}_{1,2}(\lambda)\check{R}_{3,4}(\lambda)\check{R}_{2,3}(\lambda\!+\!\delta).
\end{align}
Here, $\check{R}_{1,2}(\lambda)=P_{1,2}R_{1,2}(\lambda)$ are unitary matrices, with $R_{1,2}(\lambda)$ defined in Eq.~\eqref{eq:R-matrix}. Expanding in $\lambda$, we find
\begin{align}
    \mathds{R}_{[1,4]}(\lambda;\delta)=\mathds{1}\!+\!\ii\lambda \mathcal{H}_{[1,4]}
    +\frac{(\ii\lambda)^2}{2!}\mathcal{H}_{[1,4]}^2
    +O(\lambda^3),
    \label{eq:almost-exp-H}
\end{align}
where $\mathcal{H}_{[1,4]}=4 h_{[1,4]}\!+\!(2\!-\!\eta)\mathds{1}$. The third order in $\lambda$ cannot be expressed in terms of powers of $\mathcal{H}_{[1,4]}$ only. Therefore, up to the third order in $\lambda$, the matrix $\mathds{R}_{[1,4]}(\lambda;\delta)$ is equal to the exponential of the local Hamiltonian. As such, it can be chosen as the unitary gate in the integrable brickwork circuit that, in the Trotter-Suzuki limit, reproduces the continuous-time dynamics of the chiral spin ladder. In particular, denoting the gate as $U_{[1,4]}=\mathds{R}_{[1,4]}(\tau;\delta)$, with $\delta=-\sqrt{\eta/(1-\eta)}$, the one-step propagator of the Trotterized chiral spin ladder becomes $\mathds{U}$, reported in Eq.~\eqref{eq:brickwork-circuit}. The Trotter-Suzuki limit is then $\ee^{-\ii t H}=\lim_{n\to\infty}\mathds{U}^{n}$. 

The transfer matrix from which the circuit and its conserved quantities are generated can be identified following Ref.~\cite{paletta2025}, which deals with construction of inhomogeneous integrable-circuit architectures. Specifically, the transfer matrix reads
\begin{widetext}
\begin{align}
    \mathds{T}_{\tau,\delta}(\lambda)={\rm Tr}_a\left[\prod_{1\le x \le L/4}^\rightarrow R_{4j-3,a}\left(\lambda\!+\!\tfrac{\tau-\delta}{2}\right)R_{4j-2,a}\left(\lambda\!+\!\tfrac{\tau+\delta}{2}\right)R_{4j-1,a}\left(\lambda\!-\!\tfrac{\tau+\delta}{2}\right)R_{4j,a}\left(\lambda\!-\!\tfrac{\tau-\delta}{2}\right)\right],
\end{align}
\end{widetext}
where the $R$ matrix is given in Eq.~\eqref{eq:R-matrix}.

\section{Convergence of the TEBD simulations}
\label{app:convergence}

\begin{figure}
    \centering
    \includegraphics[width=\linewidth]{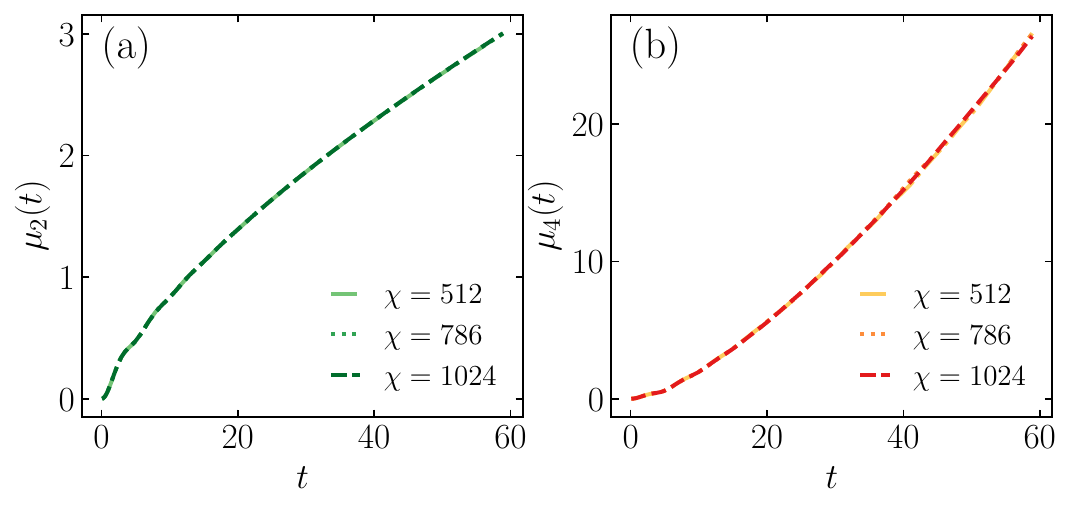}
    \caption{\textbf{Hamiltonian dynamics:} Convergence test for the second and fourth moments of the magnetization transfer for different bond dimensions. The other parameters are $L=128,\ dt=0.1, \eta=0.5$.  }
    \label{fig:convergence_ladder}
\end{figure}

\begin{figure}[b]
    \centering
    \includegraphics[width=\linewidth]{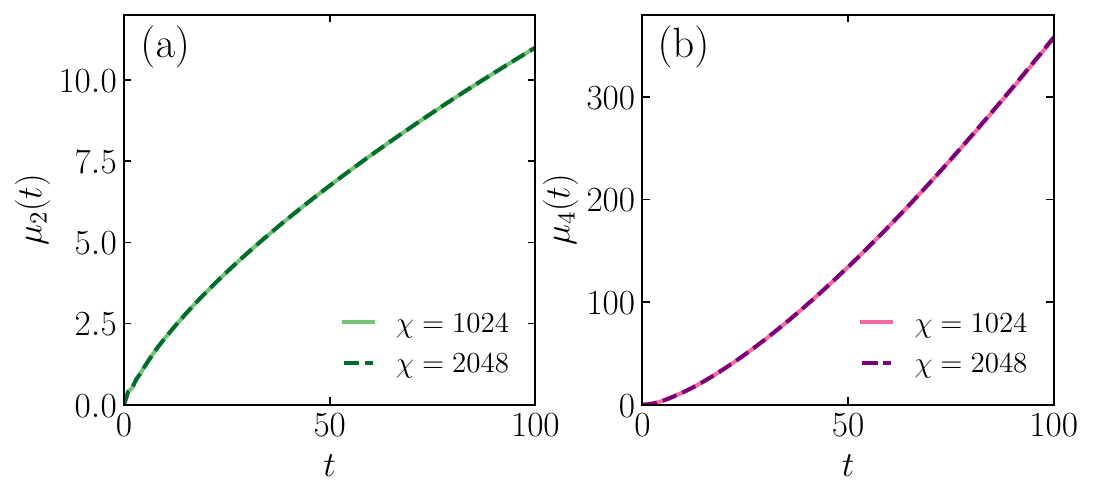}
    \caption{\textbf{Circuit dynamics:} Convergence test for the second and fourth moments of the magnetization transfer for different bond dimensions. The other parameters are $L=256,\ \tau=1, \delta=1.0$.  }
    \label{fig:convergence_circuit}
\end{figure}

In this section, we present additional simulation results to check the convergence of the data for both the Hamiltonian and the circuit dynamics. We first present the results for the chiral ladder for $\eta=0.5$ and plot the dynamics of the second $\mu_2(t)$ and fourth moment ($\mu_4(t)$) in Fig.~\ref{fig:convergence_ladder} for $L=128$ and for different bond dimensions $\chi=512,\ 786,\ 1024$ with $dt=0.1$. A similar analysis for the circuit dynamics is plotted in Fig.~\ref{fig:convergence_circuit} for $L=256$ and for two different bond dimensions $\chi = 1024,\ 2048$ with $\tau=1$ and for $\delta=1.0$.
As can be seen, both the second and the fourth moments are well converged in the bond dimension for both the continuous and discrete time evolution, and thus provide a reliable estimate for the excess kurtosis presented in the main section.

\newpage

\bibliographystyle{apsrev4-2}
\bibliography{paper.bbl}

\begin{thebibliography}{71}%
\makeatletter
\providecommand \@ifxundefined [1]{%
 \@ifx{#1\undefined}
}%
\providecommand \@ifnum [1]{%
 \ifnum #1\expandafter \@firstoftwo
 \else \expandafter \@secondoftwo
 \fi
}%
\providecommand \@ifx [1]{%
 \ifx #1\expandafter \@firstoftwo
 \else \expandafter \@secondoftwo
 \fi
}%
\providecommand \natexlab [1]{#1}%
\providecommand \enquote  [1]{``#1''}%
\providecommand \bibnamefont  [1]{#1}%
\providecommand \bibfnamefont [1]{#1}%
\providecommand \citenamefont [1]{#1}%
\providecommand \href@noop [0]{\@secondoftwo}%
\providecommand \href [0]{\begingroup \@sanitize@url \@href}%
\providecommand \@href[1]{\@@startlink{#1}\@@href}%
\providecommand \@@href[1]{\endgroup#1\@@endlink}%
\providecommand \@sanitize@url [0]{\catcode `\\12\catcode `\$12\catcode `\&12\catcode `\#12\catcode `\^12\catcode `\_12\catcode `\%12\relax}%
\providecommand \@@startlink[1]{}%
\providecommand \@@endlink[0]{}%
\providecommand \url  [0]{\begingroup\@sanitize@url \@url }%
\providecommand \@url [1]{\endgroup\@href {#1}{\urlprefix }}%
\providecommand \urlprefix  [0]{URL }%
\providecommand \Eprint [0]{\href }%
\providecommand \doibase [0]{https://doi.org/}%
\providecommand \selectlanguage [0]{\@gobble}%
\providecommand \bibinfo  [0]{\@secondoftwo}%
\providecommand \bibfield  [0]{\@secondoftwo}%
\providecommand \translation [1]{[#1]}%
\providecommand \BibitemOpen [0]{}%
\providecommand \bibitemStop [0]{}%
\providecommand \bibitemNoStop [0]{.\EOS\space}%
\providecommand \EOS [0]{\spacefactor3000\relax}%
\providecommand \BibitemShut  [1]{\csname bibitem#1\endcsname}%
\let\auto@bib@innerbib\@empty
\bibitem [{\citenamefont {Romatschke}\ and\ \citenamefont {Romatschke}(2007)}]{Romatschke2007Viscosity}%
  \BibitemOpen
  \bibfield  {author} {\bibinfo {author} {\bibfnamefont {P.}~\bibnamefont {Romatschke}}\ and\ \bibinfo {author} {\bibfnamefont {U.}~\bibnamefont {Romatschke}},\ }\href {https://doi.org/10.1103/PhysRevLett.99.172301} {\bibfield  {journal} {\bibinfo  {journal} {Phys. Rev. Lett.}\ }\textbf {\bibinfo {volume} {99}},\ \bibinfo {pages} {172301} (\bibinfo {year} {2007})}\BibitemShut {NoStop}%
\bibitem [{\citenamefont {Teaney}(2010)}]{Teaney2010Viscous}%
  \BibitemOpen
  \bibfield  {author} {\bibinfo {author} {\bibfnamefont {D.~A.}\ \bibnamefont {Teaney}},\ }\bibinfo {title} {{Viscous Hydrodynamics and the Quark Gluon Plasma}},\ in\ \href {https://doi.org/10.1142/9789814293297_0004} {\emph {\bibinfo {booktitle} {Quark-Gluon Plasma 4}}}\ (\bibinfo  {publisher} {World Scientific},\ \bibinfo {year} {2010})\ p.\ \bibinfo {pages} {207–266}\BibitemShut {NoStop}%
\bibitem [{\citenamefont {Snellings}(2011)}]{Snellings2011Elliptic}%
  \BibitemOpen
  \bibfield  {author} {\bibinfo {author} {\bibfnamefont {R.}~\bibnamefont {Snellings}},\ }\href {https://doi.org/10.1088/1367-2630/13/5/055008} {\bibfield  {journal} {\bibinfo  {journal} {New Journal of Physics}\ }\textbf {\bibinfo {volume} {13}},\ \bibinfo {pages} {055008} (\bibinfo {year} {2011})}\BibitemShut {NoStop}%
\bibitem [{\citenamefont {Busza}\ \emph {et~al.}(2018)\citenamefont {Busza}, \citenamefont {Rajagopal},\ and\ \citenamefont {van~der Schee}}]{Busza2018Heavy}%
  \BibitemOpen
  \bibfield  {author} {\bibinfo {author} {\bibfnamefont {W.}~\bibnamefont {Busza}}, \bibinfo {author} {\bibfnamefont {K.}~\bibnamefont {Rajagopal}},\ and\ \bibinfo {author} {\bibfnamefont {W.}~\bibnamefont {van~der Schee}},\ }\href {https://doi.org/https://doi.org/10.1146/annurev-nucl-101917-020852} {\bibfield  {journal} {\bibinfo  {journal} {Annual Review of Nuclear and Particle Science}\ }\textbf {\bibinfo {volume} {68}},\ \bibinfo {pages} {339} (\bibinfo {year} {2018})}\BibitemShut {NoStop}%
\bibitem [{\citenamefont {Schneider}\ \emph {et~al.}(2012)\citenamefont {Schneider}, \citenamefont {Hackerm{\"u}ller}, \citenamefont {Ronzheimer}, \citenamefont {Will}, \citenamefont {Braun}, \citenamefont {Best}, \citenamefont {Bloch}, \citenamefont {Demler}, \citenamefont {Mandt}, \citenamefont {Rasch},\ and\ \citenamefont {Rosch}}]{Schneider2012Fermionic}%
  \BibitemOpen
  \bibfield  {author} {\bibinfo {author} {\bibfnamefont {U.}~\bibnamefont {Schneider}}, \bibinfo {author} {\bibfnamefont {L.}~\bibnamefont {Hackerm{\"u}ller}}, \bibinfo {author} {\bibfnamefont {J.~P.}\ \bibnamefont {Ronzheimer}}, \bibinfo {author} {\bibfnamefont {S.}~\bibnamefont {Will}}, \bibinfo {author} {\bibfnamefont {S.}~\bibnamefont {Braun}}, \bibinfo {author} {\bibfnamefont {T.}~\bibnamefont {Best}}, \bibinfo {author} {\bibfnamefont {I.}~\bibnamefont {Bloch}}, \bibinfo {author} {\bibfnamefont {E.}~\bibnamefont {Demler}}, \bibinfo {author} {\bibfnamefont {S.}~\bibnamefont {Mandt}}, \bibinfo {author} {\bibfnamefont {D.}~\bibnamefont {Rasch}},\ and\ \bibinfo {author} {\bibfnamefont {A.}~\bibnamefont {Rosch}},\ }\href {https://doi.org/10.1038/nphys2205} {\bibfield  {journal} {\bibinfo  {journal} {Nature Physics}\ }\textbf {\bibinfo {volume} {8}},\ \bibinfo {pages} {213} (\bibinfo {year} {2012})}\BibitemShut {NoStop}%
\bibitem [{\citenamefont {Trotzky}\ \emph {et~al.}(2012)\citenamefont {Trotzky}, \citenamefont {Chen}, \citenamefont {Flesch}, \citenamefont {McCulloch}, \citenamefont {Schollw{\"o}ck}, \citenamefont {Eisert},\ and\ \citenamefont {Bloch}}]{Trotzky2012Probing}%
  \BibitemOpen
  \bibfield  {author} {\bibinfo {author} {\bibfnamefont {S.}~\bibnamefont {Trotzky}}, \bibinfo {author} {\bibfnamefont {Y.-A.}\ \bibnamefont {Chen}}, \bibinfo {author} {\bibfnamefont {A.}~\bibnamefont {Flesch}}, \bibinfo {author} {\bibfnamefont {I.~P.}\ \bibnamefont {McCulloch}}, \bibinfo {author} {\bibfnamefont {U.}~\bibnamefont {Schollw{\"o}ck}}, \bibinfo {author} {\bibfnamefont {J.}~\bibnamefont {Eisert}},\ and\ \bibinfo {author} {\bibfnamefont {I.}~\bibnamefont {Bloch}},\ }\href {https://doi.org/10.1038/nphys2232} {\bibfield  {journal} {\bibinfo  {journal} {Nature Physics}\ }\textbf {\bibinfo {volume} {8}},\ \bibinfo {pages} {325} (\bibinfo {year} {2012})}\BibitemShut {NoStop}%
\bibitem [{\citenamefont {Hild}\ \emph {et~al.}(2014)\citenamefont {Hild}, \citenamefont {Fukuhara}, \citenamefont {Schau\ss{}}, \citenamefont {Zeiher}, \citenamefont {Knap}, \citenamefont {Demler}, \citenamefont {Bloch},\ and\ \citenamefont {Gross}}]{Fukuhara2014Far-from-equilibrium}%
  \BibitemOpen
  \bibfield  {author} {\bibinfo {author} {\bibfnamefont {S.}~\bibnamefont {Hild}}, \bibinfo {author} {\bibfnamefont {T.}~\bibnamefont {Fukuhara}}, \bibinfo {author} {\bibfnamefont {P.}~\bibnamefont {Schau\ss{}}}, \bibinfo {author} {\bibfnamefont {J.}~\bibnamefont {Zeiher}}, \bibinfo {author} {\bibfnamefont {M.}~\bibnamefont {Knap}}, \bibinfo {author} {\bibfnamefont {E.}~\bibnamefont {Demler}}, \bibinfo {author} {\bibfnamefont {I.}~\bibnamefont {Bloch}},\ and\ \bibinfo {author} {\bibfnamefont {C.}~\bibnamefont {Gross}},\ }\href {https://doi.org/10.1103/PhysRevLett.113.147205} {\bibfield  {journal} {\bibinfo  {journal} {Phys. Rev. Lett.}\ }\textbf {\bibinfo {volume} {113}},\ \bibinfo {pages} {147205} (\bibinfo {year} {2014})}\BibitemShut {NoStop}%
\bibitem [{\citenamefont {Georgescu}\ \emph {et~al.}(2014)\citenamefont {Georgescu}, \citenamefont {Ashhab},\ and\ \citenamefont {Nori}}]{Georgescu2014Quantum}%
  \BibitemOpen
  \bibfield  {author} {\bibinfo {author} {\bibfnamefont {I.~M.}\ \bibnamefont {Georgescu}}, \bibinfo {author} {\bibfnamefont {S.}~\bibnamefont {Ashhab}},\ and\ \bibinfo {author} {\bibfnamefont {F.}~\bibnamefont {Nori}},\ }\href {https://doi.org/10.1103/RevModPhys.86.153} {\bibfield  {journal} {\bibinfo  {journal} {Rev. Mod. Phys.}\ }\textbf {\bibinfo {volume} {86}},\ \bibinfo {pages} {153} (\bibinfo {year} {2014})}\BibitemShut {NoStop}%
\bibitem [{\citenamefont {Brown}\ \emph {et~al.}(2015)\citenamefont {Brown}, \citenamefont {Wyllie}, \citenamefont {Koller}, \citenamefont {Goldschmidt}, \citenamefont {Foss-Feig},\ and\ \citenamefont {Porto}}]{Brown2015Two-dimensional}%
  \BibitemOpen
  \bibfield  {author} {\bibinfo {author} {\bibfnamefont {R.~C.}\ \bibnamefont {Brown}}, \bibinfo {author} {\bibfnamefont {R.}~\bibnamefont {Wyllie}}, \bibinfo {author} {\bibfnamefont {S.~B.}\ \bibnamefont {Koller}}, \bibinfo {author} {\bibfnamefont {E.~A.}\ \bibnamefont {Goldschmidt}}, \bibinfo {author} {\bibfnamefont {M.}~\bibnamefont {Foss-Feig}},\ and\ \bibinfo {author} {\bibfnamefont {J.~V.}\ \bibnamefont {Porto}},\ }\href {https://doi.org/10.1126/science.aaa1385} {\bibfield  {journal} {\bibinfo  {journal} {Science}\ }\textbf {\bibinfo {volume} {348}},\ \bibinfo {pages} {540} (\bibinfo {year} {2015})}\BibitemShut {NoStop}%
\bibitem [{\citenamefont {Chien}\ \emph {et~al.}(2015)\citenamefont {Chien}, \citenamefont {Peotta},\ and\ \citenamefont {Di~Ventra}}]{Chien2015Quantum}%
  \BibitemOpen
  \bibfield  {author} {\bibinfo {author} {\bibfnamefont {C.-C.}\ \bibnamefont {Chien}}, \bibinfo {author} {\bibfnamefont {S.}~\bibnamefont {Peotta}},\ and\ \bibinfo {author} {\bibfnamefont {M.}~\bibnamefont {Di~Ventra}},\ }\href {https://doi.org/10.1038/nphys3531} {\bibfield  {journal} {\bibinfo  {journal} {Nature Physics}\ }\textbf {\bibinfo {volume} {11}},\ \bibinfo {pages} {998} (\bibinfo {year} {2015})}\BibitemShut {NoStop}%
\bibitem [{\citenamefont {Gross}\ and\ \citenamefont {Bloch}(2017)}]{Gross2017Quantum}%
  \BibitemOpen
  \bibfield  {author} {\bibinfo {author} {\bibfnamefont {C.}~\bibnamefont {Gross}}\ and\ \bibinfo {author} {\bibfnamefont {I.}~\bibnamefont {Bloch}},\ }\href {https://doi.org/10.1126/science.aal3837} {\bibfield  {journal} {\bibinfo  {journal} {Science}\ }\textbf {\bibinfo {volume} {357}},\ \bibinfo {pages} {995} (\bibinfo {year} {2017})}\BibitemShut {NoStop}%
\bibitem [{\citenamefont {Nichols}\ \emph {et~al.}(2019)\citenamefont {Nichols}, \citenamefont {Cheuk}, \citenamefont {Okan}, \citenamefont {Hartke}, \citenamefont {Mendez}, \citenamefont {Senthil}, \citenamefont {Khatami}, \citenamefont {Zhang},\ and\ \citenamefont {Zwierlein}}]{Nichols2019Spin}%
  \BibitemOpen
  \bibfield  {author} {\bibinfo {author} {\bibfnamefont {M.~A.}\ \bibnamefont {Nichols}}, \bibinfo {author} {\bibfnamefont {L.~W.}\ \bibnamefont {Cheuk}}, \bibinfo {author} {\bibfnamefont {M.}~\bibnamefont {Okan}}, \bibinfo {author} {\bibfnamefont {T.~R.}\ \bibnamefont {Hartke}}, \bibinfo {author} {\bibfnamefont {E.}~\bibnamefont {Mendez}}, \bibinfo {author} {\bibfnamefont {T.}~\bibnamefont {Senthil}}, \bibinfo {author} {\bibfnamefont {E.}~\bibnamefont {Khatami}}, \bibinfo {author} {\bibfnamefont {H.}~\bibnamefont {Zhang}},\ and\ \bibinfo {author} {\bibfnamefont {M.~W.}\ \bibnamefont {Zwierlein}},\ }\href {https://doi.org/10.1126/science.aat4387} {\bibfield  {journal} {\bibinfo  {journal} {Science}\ }\textbf {\bibinfo {volume} {363}},\ \bibinfo {pages} {383} (\bibinfo {year} {2019})}\BibitemShut {NoStop}%
\bibitem [{\citenamefont {Brown}\ \emph {et~al.}(2019)\citenamefont {Brown}, \citenamefont {Mitra}, \citenamefont {Guardado-Sanchez}, \citenamefont {Nourafkan}, \citenamefont {Reymbaut}, \citenamefont {H{\'e}bert}, \citenamefont {Bergeron}, \citenamefont {Tremblay}, \citenamefont {Kokalj}, \citenamefont {Huse}, \citenamefont {Schau{\ss}},\ and\ \citenamefont {Bakr}}]{Brown2019Bad}%
  \BibitemOpen
  \bibfield  {author} {\bibinfo {author} {\bibfnamefont {P.~T.}\ \bibnamefont {Brown}}, \bibinfo {author} {\bibfnamefont {D.}~\bibnamefont {Mitra}}, \bibinfo {author} {\bibfnamefont {E.}~\bibnamefont {Guardado-Sanchez}}, \bibinfo {author} {\bibfnamefont {R.}~\bibnamefont {Nourafkan}}, \bibinfo {author} {\bibfnamefont {A.}~\bibnamefont {Reymbaut}}, \bibinfo {author} {\bibfnamefont {C.-D.}\ \bibnamefont {H{\'e}bert}}, \bibinfo {author} {\bibfnamefont {S.}~\bibnamefont {Bergeron}}, \bibinfo {author} {\bibfnamefont {A.-M.~S.}\ \bibnamefont {Tremblay}}, \bibinfo {author} {\bibfnamefont {J.}~\bibnamefont {Kokalj}}, \bibinfo {author} {\bibfnamefont {D.~A.}\ \bibnamefont {Huse}}, \bibinfo {author} {\bibfnamefont {P.}~\bibnamefont {Schau{\ss}}},\ and\ \bibinfo {author} {\bibfnamefont {W.~S.}\ \bibnamefont {Bakr}},\ }\href {https://doi.org/10.1126/science.aat4134} {\bibfield  {journal} {\bibinfo  {journal} {Science}\ }\textbf {\bibinfo {volume} {363}},\ \bibinfo {pages} {379} (\bibinfo {year} {2019})}\BibitemShut
  {NoStop}%
\bibitem [{\citenamefont {\ifmmode \check{Z}\else \v{Z}\fi{}nidari\ifmmode~\check{c}\else \v{c}\fi{}}(2011)}]{ZnidarichSpin}%
  \BibitemOpen
  \bibfield  {author} {\bibinfo {author} {\bibfnamefont {M.}~\bibnamefont {\ifmmode \check{Z}\else \v{Z}\fi{}nidari\ifmmode~\check{c}\else \v{c}\fi{}}},\ }\href {https://doi.org/10.1103/PhysRevLett.106.220601} {\bibfield  {journal} {\bibinfo  {journal} {Phys. Rev. Lett.}\ }\textbf {\bibinfo {volume} {106}},\ \bibinfo {pages} {220601} (\bibinfo {year} {2011})}\BibitemShut {NoStop}%
\bibitem [{\citenamefont {Ljubotina}\ \emph {et~al.}(2019)\citenamefont {Ljubotina}, \citenamefont {\ifmmode \check{Z}\else \v{Z}\fi{}nidari\ifmmode~\check{c}\else \v{c}\fi{}},\ and\ \citenamefont {Prosen}}]{ljubotina2019kardar}%
  \BibitemOpen
  \bibfield  {author} {\bibinfo {author} {\bibfnamefont {M.}~\bibnamefont {Ljubotina}}, \bibinfo {author} {\bibfnamefont {M.}~\bibnamefont {\ifmmode \check{Z}\else \v{Z}\fi{}nidari\ifmmode~\check{c}\else \v{c}\fi{}}},\ and\ \bibinfo {author} {\bibfnamefont {T.~c.~v.}\ \bibnamefont {Prosen}},\ }\href {https://doi.org/10.1103/PhysRevLett.122.210602} {\bibfield  {journal} {\bibinfo  {journal} {Phys. Rev. Lett.}\ }\textbf {\bibinfo {volume} {122}},\ \bibinfo {pages} {210602} (\bibinfo {year} {2019})}\BibitemShut {NoStop}%
\bibitem [{\citenamefont {Dupont}\ and\ \citenamefont {Moore}(2020)}]{DupontMoore}%
  \BibitemOpen
  \bibfield  {author} {\bibinfo {author} {\bibfnamefont {M.}~\bibnamefont {Dupont}}\ and\ \bibinfo {author} {\bibfnamefont {J.~E.}\ \bibnamefont {Moore}},\ }\href {https://doi.org/10.1103/PhysRevB.101.121106} {\bibfield  {journal} {\bibinfo  {journal} {Phys. Rev. B}\ }\textbf {\bibinfo {volume} {101}},\ \bibinfo {pages} {121106} (\bibinfo {year} {2020})}\BibitemShut {NoStop}%
\bibitem [{\citenamefont {Bar~Lev}\ \emph {et~al.}(2015)\citenamefont {Bar~Lev}, \citenamefont {Cohen},\ and\ \citenamefont {Reichman}}]{barlev2015absence}%
  \BibitemOpen
  \bibfield  {author} {\bibinfo {author} {\bibfnamefont {Y.}~\bibnamefont {Bar~Lev}}, \bibinfo {author} {\bibfnamefont {G.}~\bibnamefont {Cohen}},\ and\ \bibinfo {author} {\bibfnamefont {D.~R.}\ \bibnamefont {Reichman}},\ }\href {https://doi.org/10.1103/PhysRevLett.114.100601} {\bibfield  {journal} {\bibinfo  {journal} {Phys. Rev. Lett.}\ }\textbf {\bibinfo {volume} {114}},\ \bibinfo {pages} {100601} (\bibinfo {year} {2015})}\BibitemShut {NoStop}%
\bibitem [{\citenamefont {Agarwal}\ \emph {et~al.}(2015)\citenamefont {Agarwal}, \citenamefont {Gopalakrishnan}, \citenamefont {Knap}, \citenamefont {M\"uller},\ and\ \citenamefont {Demler}}]{agarwal2015anomalous}%
  \BibitemOpen
  \bibfield  {author} {\bibinfo {author} {\bibfnamefont {K.}~\bibnamefont {Agarwal}}, \bibinfo {author} {\bibfnamefont {S.}~\bibnamefont {Gopalakrishnan}}, \bibinfo {author} {\bibfnamefont {M.}~\bibnamefont {Knap}}, \bibinfo {author} {\bibfnamefont {M.}~\bibnamefont {M\"uller}},\ and\ \bibinfo {author} {\bibfnamefont {E.}~\bibnamefont {Demler}},\ }\href {https://doi.org/10.1103/PhysRevLett.114.160401} {\bibfield  {journal} {\bibinfo  {journal} {Phys. Rev. Lett.}\ }\textbf {\bibinfo {volume} {114}},\ \bibinfo {pages} {160401} (\bibinfo {year} {2015})}\BibitemShut {NoStop}%
\bibitem [{\citenamefont {Bar~Lev}\ \emph {et~al.}(2017)\citenamefont {Bar~Lev}, \citenamefont {Kennes}, \citenamefont {Klöckner}, \citenamefont {Reichman},\ and\ \citenamefont {Karrasch}}]{Bar_Lev_2017}%
  \BibitemOpen
  \bibfield  {author} {\bibinfo {author} {\bibfnamefont {Y.}~\bibnamefont {Bar~Lev}}, \bibinfo {author} {\bibfnamefont {D.~M.}\ \bibnamefont {Kennes}}, \bibinfo {author} {\bibfnamefont {C.}~\bibnamefont {Klöckner}}, \bibinfo {author} {\bibfnamefont {D.~R.}\ \bibnamefont {Reichman}},\ and\ \bibinfo {author} {\bibfnamefont {C.}~\bibnamefont {Karrasch}},\ }\href {https://doi.org/10.1209/0295-5075/119/37003} {\bibfield  {journal} {\bibinfo  {journal} {Europhysics Letters}\ }\textbf {\bibinfo {volume} {119}},\ \bibinfo {pages} {37003} (\bibinfo {year} {2017})}\BibitemShut {NoStop}%
\bibitem [{\citenamefont {Feldmeier}\ \emph {et~al.}(2020)\citenamefont {Feldmeier}, \citenamefont {Sala}, \citenamefont {De~Tomasi}, \citenamefont {Pollmann},\ and\ \citenamefont {Knap}}]{feldmeier2020anomalous}%
  \BibitemOpen
  \bibfield  {author} {\bibinfo {author} {\bibfnamefont {J.}~\bibnamefont {Feldmeier}}, \bibinfo {author} {\bibfnamefont {P.}~\bibnamefont {Sala}}, \bibinfo {author} {\bibfnamefont {G.}~\bibnamefont {De~Tomasi}}, \bibinfo {author} {\bibfnamefont {F.}~\bibnamefont {Pollmann}},\ and\ \bibinfo {author} {\bibfnamefont {M.}~\bibnamefont {Knap}},\ }\href {https://doi.org/10.1103/PhysRevLett.125.245303} {\bibfield  {journal} {\bibinfo  {journal} {Phys. Rev. Lett.}\ }\textbf {\bibinfo {volume} {125}},\ \bibinfo {pages} {245303} (\bibinfo {year} {2020})}\BibitemShut {NoStop}%
\bibitem [{\citenamefont {Morningstar}\ \emph {et~al.}(2020)\citenamefont {Morningstar}, \citenamefont {Khemani},\ and\ \citenamefont {Huse}}]{morningstar2020kinetically}%
  \BibitemOpen
  \bibfield  {author} {\bibinfo {author} {\bibfnamefont {A.}~\bibnamefont {Morningstar}}, \bibinfo {author} {\bibfnamefont {V.}~\bibnamefont {Khemani}},\ and\ \bibinfo {author} {\bibfnamefont {D.~A.}\ \bibnamefont {Huse}},\ }\href {https://doi.org/10.1103/PhysRevB.101.214205} {\bibfield  {journal} {\bibinfo  {journal} {Phys. Rev. B}\ }\textbf {\bibinfo {volume} {101}},\ \bibinfo {pages} {214205} (\bibinfo {year} {2020})}\BibitemShut {NoStop}%
\bibitem [{\citenamefont {Moudgalya}\ \emph {et~al.}(2021)\citenamefont {Moudgalya}, \citenamefont {Prem}, \citenamefont {Huse},\ and\ \citenamefont {Chan}}]{moudgalya2021spectral}%
  \BibitemOpen
  \bibfield  {author} {\bibinfo {author} {\bibfnamefont {S.}~\bibnamefont {Moudgalya}}, \bibinfo {author} {\bibfnamefont {A.}~\bibnamefont {Prem}}, \bibinfo {author} {\bibfnamefont {D.~A.}\ \bibnamefont {Huse}},\ and\ \bibinfo {author} {\bibfnamefont {A.}~\bibnamefont {Chan}},\ }\href {https://doi.org/10.1103/PhysRevResearch.3.023176} {\bibfield  {journal} {\bibinfo  {journal} {Phys. Rev. Res.}\ }\textbf {\bibinfo {volume} {3}},\ \bibinfo {pages} {023176} (\bibinfo {year} {2021})}\BibitemShut {NoStop}%
\bibitem [{\citenamefont {Gromov}\ \emph {et~al.}(2020)\citenamefont {Gromov}, \citenamefont {Lucas},\ and\ \citenamefont {Nandkishore}}]{gromov2020fracton}%
  \BibitemOpen
  \bibfield  {author} {\bibinfo {author} {\bibfnamefont {A.}~\bibnamefont {Gromov}}, \bibinfo {author} {\bibfnamefont {A.}~\bibnamefont {Lucas}},\ and\ \bibinfo {author} {\bibfnamefont {R.~M.}\ \bibnamefont {Nandkishore}},\ }\href {https://doi.org/10.1103/PhysRevResearch.2.033124} {\bibfield  {journal} {\bibinfo  {journal} {Phys. Rev. Res.}\ }\textbf {\bibinfo {volume} {2}},\ \bibinfo {pages} {033124} (\bibinfo {year} {2020})}\BibitemShut {NoStop}%
\bibitem [{\citenamefont {Iaconis}\ \emph {et~al.}(2021)\citenamefont {Iaconis}, \citenamefont {Lucas},\ and\ \citenamefont {Nandkishore}}]{iaconis2021multipole}%
  \BibitemOpen
  \bibfield  {author} {\bibinfo {author} {\bibfnamefont {J.}~\bibnamefont {Iaconis}}, \bibinfo {author} {\bibfnamefont {A.}~\bibnamefont {Lucas}},\ and\ \bibinfo {author} {\bibfnamefont {R.}~\bibnamefont {Nandkishore}},\ }\href {https://doi.org/10.1103/PhysRevE.103.022142} {\bibfield  {journal} {\bibinfo  {journal} {Phys. Rev. E}\ }\textbf {\bibinfo {volume} {103}},\ \bibinfo {pages} {022142} (\bibinfo {year} {2021})}\BibitemShut {NoStop}%
\bibitem [{\citenamefont {\ifmmode \check{Z}\else \v{Z}\fi{}nidari\ifmmode~\check{c}\else \v{c}\fi{}}(2024)}]{znidaric2024superdiffusive}%
  \BibitemOpen
  \bibfield  {author} {\bibinfo {author} {\bibfnamefont {M.}~\bibnamefont {\ifmmode \check{Z}\else \v{Z}\fi{}nidari\ifmmode~\check{c}\else \v{c}\fi{}}},\ }\href {https://doi.org/10.1103/PhysRevB.109.075105} {\bibfield  {journal} {\bibinfo  {journal} {Phys. Rev. B}\ }\textbf {\bibinfo {volume} {109}},\ \bibinfo {pages} {075105} (\bibinfo {year} {2024})}\BibitemShut {NoStop}%
\bibitem [{\citenamefont {Ljubotina}\ \emph {et~al.}(2023)\citenamefont {Ljubotina}, \citenamefont {Desaules}, \citenamefont {Serbyn},\ and\ \citenamefont {Papi\ifmmode~\acute{c}\else \'{c}\fi{}}}]{ljubotina2023superdiffusive}%
  \BibitemOpen
  \bibfield  {author} {\bibinfo {author} {\bibfnamefont {M.}~\bibnamefont {Ljubotina}}, \bibinfo {author} {\bibfnamefont {J.-Y.}\ \bibnamefont {Desaules}}, \bibinfo {author} {\bibfnamefont {M.}~\bibnamefont {Serbyn}},\ and\ \bibinfo {author} {\bibfnamefont {Z.}~\bibnamefont {Papi\ifmmode~\acute{c}\else \'{c}\fi{}}},\ }\href {https://doi.org/10.1103/PhysRevX.13.011033} {\bibfield  {journal} {\bibinfo  {journal} {Phys. Rev. X}\ }\textbf {\bibinfo {volume} {13}},\ \bibinfo {pages} {011033} (\bibinfo {year} {2023})}\BibitemShut {NoStop}%
\bibitem [{\citenamefont {Wang}\ \emph {et~al.}(2024)\citenamefont {Wang}, \citenamefont {Fang},\ and\ \citenamefont {Ren}}]{yupeng2024superdiffusive}%
  \BibitemOpen
  \bibfield  {author} {\bibinfo {author} {\bibfnamefont {Y.-P.}\ \bibnamefont {Wang}}, \bibinfo {author} {\bibfnamefont {C.}~\bibnamefont {Fang}},\ and\ \bibinfo {author} {\bibfnamefont {J.}~\bibnamefont {Ren}},\ }\href {https://doi.org/10.21468/SciPostPhys.17.6.150} {\bibfield  {journal} {\bibinfo  {journal} {SciPost Phys.}\ }\textbf {\bibinfo {volume} {17}},\ \bibinfo {pages} {150} (\bibinfo {year} {2024})}\BibitemShut {NoStop}%
\bibitem [{\citenamefont {Bhakuni}\ \emph {et~al.}(2025)\citenamefont {Bhakuni}, \citenamefont {Verdel}, \citenamefont {Desaules}, \citenamefont {Serbyn}, \citenamefont {Ljubotina},\ and\ \citenamefont {Dalmonte}}]{bhakuni2025anomalously}%
  \BibitemOpen
  \bibfield  {author} {\bibinfo {author} {\bibfnamefont {D.~S.}\ \bibnamefont {Bhakuni}}, \bibinfo {author} {\bibfnamefont {R.}~\bibnamefont {Verdel}}, \bibinfo {author} {\bibfnamefont {J.-Y.}\ \bibnamefont {Desaules}}, \bibinfo {author} {\bibfnamefont {M.}~\bibnamefont {Serbyn}}, \bibinfo {author} {\bibfnamefont {M.}~\bibnamefont {Ljubotina}},\ and\ \bibinfo {author} {\bibfnamefont {M.}~\bibnamefont {Dalmonte}},\ }\href {https://arxiv.org/abs/2509.08889} {\bibinfo {title} {Anomalously fast transport in non-integrable lattice gauge theories}} (\bibinfo {year} {2025}),\ \Eprint {https://arxiv.org/abs/2509.08889} {arXiv:2509.08889} \BibitemShut {NoStop}%
\bibitem [{\citenamefont {Moca}\ \emph {et~al.}(2025)\citenamefont {Moca}, \citenamefont {D\'{o}ra}, \citenamefont {Sticlet}, \citenamefont {Valli}, \citenamefont {Prosen},\ and\ \citenamefont {Zar\'{a}nd}}]{moca2025dynamicscalingfamilyvicsekuniversality}%
  \BibitemOpen
  \bibfield  {author} {\bibinfo {author} {\bibfnamefont {C.~P.}\ \bibnamefont {Moca}}, \bibinfo {author} {\bibfnamefont {B.}~\bibnamefont {D\'{o}ra}}, \bibinfo {author} {\bibfnamefont {D.}~\bibnamefont {Sticlet}}, \bibinfo {author} {\bibfnamefont {A.}~\bibnamefont {Valli}}, \bibinfo {author} {\bibfnamefont {T.}~\bibnamefont {Prosen}},\ and\ \bibinfo {author} {\bibfnamefont {G.}~\bibnamefont {Zar\'{a}nd}},\ }\href {https://arxiv.org/abs/2503.21454} {\bibinfo {title} {{Dynamic scaling and Family-Vicsek universality in SU(N) quantum spin chains}}} (\bibinfo {year} {2025}),\ \Eprint {https://arxiv.org/abs/2503.21454} {arXiv:2503.21454} \BibitemShut {NoStop}%
\bibitem [{\citenamefont {Scheie}\ \emph {et~al.}(2021)\citenamefont {Scheie}, \citenamefont {Sherman}, \citenamefont {Dupont}, \citenamefont {Nagler}, \citenamefont {Stone}, \citenamefont {Granroth}, \citenamefont {Moore},\ and\ \citenamefont {Tennant}}]{scheie2021detection}%
  \BibitemOpen
  \bibfield  {author} {\bibinfo {author} {\bibfnamefont {A.}~\bibnamefont {Scheie}}, \bibinfo {author} {\bibfnamefont {N.}~\bibnamefont {Sherman}}, \bibinfo {author} {\bibfnamefont {M.}~\bibnamefont {Dupont}}, \bibinfo {author} {\bibfnamefont {S.}~\bibnamefont {Nagler}}, \bibinfo {author} {\bibfnamefont {M.}~\bibnamefont {Stone}}, \bibinfo {author} {\bibfnamefont {G.}~\bibnamefont {Granroth}}, \bibinfo {author} {\bibfnamefont {J.}~\bibnamefont {Moore}},\ and\ \bibinfo {author} {\bibfnamefont {D.}~\bibnamefont {Tennant}},\ }\href@noop {} {\bibfield  {journal} {\bibinfo  {journal} {Nature Physics}\ }\textbf {\bibinfo {volume} {17}},\ \bibinfo {pages} {726} (\bibinfo {year} {2021})}\BibitemShut {NoStop}%
\bibitem [{\citenamefont {Wei}\ \emph {et~al.}(2022)\citenamefont {Wei}, \citenamefont {Rubio-Abadal}, \citenamefont {Ye}, \citenamefont {Machado}, \citenamefont {Kemp}, \citenamefont {Srakaew}, \citenamefont {Hollerith}, \citenamefont {Rui}, \citenamefont {Gopalakrishnan}, \citenamefont {Yao} \emph {et~al.}}]{wei2022quantum}%
  \BibitemOpen
  \bibfield  {author} {\bibinfo {author} {\bibfnamefont {D.}~\bibnamefont {Wei}}, \bibinfo {author} {\bibfnamefont {A.}~\bibnamefont {Rubio-Abadal}}, \bibinfo {author} {\bibfnamefont {B.}~\bibnamefont {Ye}}, \bibinfo {author} {\bibfnamefont {F.}~\bibnamefont {Machado}}, \bibinfo {author} {\bibfnamefont {J.}~\bibnamefont {Kemp}}, \bibinfo {author} {\bibfnamefont {K.}~\bibnamefont {Srakaew}}, \bibinfo {author} {\bibfnamefont {S.}~\bibnamefont {Hollerith}}, \bibinfo {author} {\bibfnamefont {J.}~\bibnamefont {Rui}}, \bibinfo {author} {\bibfnamefont {S.}~\bibnamefont {Gopalakrishnan}}, \bibinfo {author} {\bibfnamefont {N.~Y.}\ \bibnamefont {Yao}}, \emph {et~al.},\ }\href@noop {} {\bibfield  {journal} {\bibinfo  {journal} {Science}\ }\textbf {\bibinfo {volume} {376}},\ \bibinfo {pages} {716} (\bibinfo {year} {2022})}\BibitemShut {NoStop}%
\bibitem [{\citenamefont {Rosenberg}\ \emph {et~al.}(2024)\citenamefont {Rosenberg}, \citenamefont {Andersen}, \citenamefont {Samajdar}, \citenamefont {Petukhov}, \citenamefont {Hoke}, \citenamefont {Abanin}, \citenamefont {Bengtsson}, \citenamefont {Drozdov}, \citenamefont {Erickson}, \citenamefont {Klimov} \emph {et~al.}}]{rosenberg2024dynamics}%
  \BibitemOpen
  \bibfield  {author} {\bibinfo {author} {\bibfnamefont {E.}~\bibnamefont {Rosenberg}}, \bibinfo {author} {\bibfnamefont {T.}~\bibnamefont {Andersen}}, \bibinfo {author} {\bibfnamefont {R.}~\bibnamefont {Samajdar}}, \bibinfo {author} {\bibfnamefont {A.}~\bibnamefont {Petukhov}}, \bibinfo {author} {\bibfnamefont {J.}~\bibnamefont {Hoke}}, \bibinfo {author} {\bibfnamefont {D.}~\bibnamefont {Abanin}}, \bibinfo {author} {\bibfnamefont {A.}~\bibnamefont {Bengtsson}}, \bibinfo {author} {\bibfnamefont {I.}~\bibnamefont {Drozdov}}, \bibinfo {author} {\bibfnamefont {C.}~\bibnamefont {Erickson}}, \bibinfo {author} {\bibfnamefont {P.}~\bibnamefont {Klimov}}, \emph {et~al.},\ }\href@noop {} {\bibfield  {journal} {\bibinfo  {journal} {Science}\ }\textbf {\bibinfo {volume} {384}},\ \bibinfo {pages} {48} (\bibinfo {year} {2024})}\BibitemShut {NoStop}%
\bibitem [{\citenamefont {Bulchandani}\ \emph {et~al.}(2021)\citenamefont {Bulchandani}, \citenamefont {Gopalakrishnan},\ and\ \citenamefont {Ilievski}}]{Bulchandani_2021}%
  \BibitemOpen
  \bibfield  {author} {\bibinfo {author} {\bibfnamefont {V.~B.}\ \bibnamefont {Bulchandani}}, \bibinfo {author} {\bibfnamefont {S.}~\bibnamefont {Gopalakrishnan}},\ and\ \bibinfo {author} {\bibfnamefont {E.}~\bibnamefont {Ilievski}},\ }\href {https://doi.org/10.1088/1742-5468/ac12c7} {\bibfield  {journal} {\bibinfo  {journal} {Journal of Statistical Mechanics: Theory and Experiment}\ }\textbf {\bibinfo {volume} {2021}},\ \bibinfo {pages} {084001} (\bibinfo {year} {2021})}\BibitemShut {NoStop}%
\bibitem [{\citenamefont {Gopalakrishnan}\ and\ \citenamefont {Vasseur}(2023)}]{Gopalakrishnan_hot}%
  \BibitemOpen
  \bibfield  {author} {\bibinfo {author} {\bibfnamefont {S.}~\bibnamefont {Gopalakrishnan}}\ and\ \bibinfo {author} {\bibfnamefont {R.}~\bibnamefont {Vasseur}},\ }\href {https://doi.org/10.1088/1361-6633/acb36e} {\bibfield  {journal} {\bibinfo  {journal} {Reports on Progress in Physics}\ }\textbf {\bibinfo {volume} {86}},\ \bibinfo {pages} {036502} (\bibinfo {year} {2023})}\BibitemShut {NoStop}%
\bibitem [{\citenamefont {Ilievski}\ \emph {et~al.}(2018)\citenamefont {Ilievski}, \citenamefont {De~Nardis}, \citenamefont {Medenjak},\ and\ \citenamefont {Prosen}}]{ilievski2018}%
  \BibitemOpen
  \bibfield  {author} {\bibinfo {author} {\bibfnamefont {E.}~\bibnamefont {Ilievski}}, \bibinfo {author} {\bibfnamefont {J.}~\bibnamefont {De~Nardis}}, \bibinfo {author} {\bibfnamefont {M.}~\bibnamefont {Medenjak}},\ and\ \bibinfo {author} {\bibfnamefont {T.}~\bibnamefont {Prosen}},\ }\href {https://doi.org/10.1103/PhysRevLett.121.230602} {\bibfield  {journal} {\bibinfo  {journal} {Phys. Rev. Lett.}\ }\textbf {\bibinfo {volume} {121}},\ \bibinfo {pages} {230602} (\bibinfo {year} {2018})}\BibitemShut {NoStop}%
\bibitem [{\citenamefont {Ilievski}\ \emph {et~al.}(2021)\citenamefont {Ilievski}, \citenamefont {De~Nardis}, \citenamefont {Gopalakrishnan}, \citenamefont {Vasseur},\ and\ \citenamefont {Ware}}]{ilievski2021superuniversality}%
  \BibitemOpen
  \bibfield  {author} {\bibinfo {author} {\bibfnamefont {E.}~\bibnamefont {Ilievski}}, \bibinfo {author} {\bibfnamefont {J.}~\bibnamefont {De~Nardis}}, \bibinfo {author} {\bibfnamefont {S.}~\bibnamefont {Gopalakrishnan}}, \bibinfo {author} {\bibfnamefont {R.}~\bibnamefont {Vasseur}},\ and\ \bibinfo {author} {\bibfnamefont {B.}~\bibnamefont {Ware}},\ }\href@noop {} {\bibfield  {journal} {\bibinfo  {journal} {Physical Review X}\ }\textbf {\bibinfo {volume} {11}},\ \bibinfo {pages} {031023} (\bibinfo {year} {2021})}\BibitemShut {NoStop}%
\bibitem [{\citenamefont {Gopalakrishnan}\ and\ \citenamefont {Vasseur}(2019)}]{GopalakrishnanKineticTheory}%
  \BibitemOpen
  \bibfield  {author} {\bibinfo {author} {\bibfnamefont {S.}~\bibnamefont {Gopalakrishnan}}\ and\ \bibinfo {author} {\bibfnamefont {R.}~\bibnamefont {Vasseur}},\ }\href {https://doi.org/10.1103/PhysRevLett.122.127202} {\bibfield  {journal} {\bibinfo  {journal} {Phys. Rev. Lett.}\ }\textbf {\bibinfo {volume} {122}},\ \bibinfo {pages} {127202} (\bibinfo {year} {2019})}\BibitemShut {NoStop}%
\bibitem [{\citenamefont {Takeuchi}\ \emph {et~al.}(2025)\citenamefont {Takeuchi}, \citenamefont {Takasan}, \citenamefont {Busani}, \citenamefont {Ferrari}, \citenamefont {Vasseur},\ and\ \citenamefont {De~Nardis}}]{TakeuchiPartialKPZ}%
  \BibitemOpen
  \bibfield  {author} {\bibinfo {author} {\bibfnamefont {K.~A.}\ \bibnamefont {Takeuchi}}, \bibinfo {author} {\bibfnamefont {K.}~\bibnamefont {Takasan}}, \bibinfo {author} {\bibfnamefont {O.}~\bibnamefont {Busani}}, \bibinfo {author} {\bibfnamefont {P.~L.}\ \bibnamefont {Ferrari}}, \bibinfo {author} {\bibfnamefont {R.}~\bibnamefont {Vasseur}},\ and\ \bibinfo {author} {\bibfnamefont {J.}~\bibnamefont {De~Nardis}},\ }\href {https://doi.org/10.1103/PhysRevLett.134.097104} {\bibfield  {journal} {\bibinfo  {journal} {Phys. Rev. Lett.}\ }\textbf {\bibinfo {volume} {134}},\ \bibinfo {pages} {097104} (\bibinfo {year} {2025})}\BibitemShut {NoStop}%
\bibitem [{\citenamefont {Das}\ \emph {et~al.}(2019)\citenamefont {Das}, \citenamefont {Kulkarni}, \citenamefont {Spohn},\ and\ \citenamefont {Dhar}}]{DasLandauLifshitz}%
  \BibitemOpen
  \bibfield  {author} {\bibinfo {author} {\bibfnamefont {A.}~\bibnamefont {Das}}, \bibinfo {author} {\bibfnamefont {M.}~\bibnamefont {Kulkarni}}, \bibinfo {author} {\bibfnamefont {H.}~\bibnamefont {Spohn}},\ and\ \bibinfo {author} {\bibfnamefont {A.}~\bibnamefont {Dhar}},\ }\href {https://doi.org/10.1103/PhysRevE.100.042116} {\bibfield  {journal} {\bibinfo  {journal} {Phys. Rev. E}\ }\textbf {\bibinfo {volume} {100}},\ \bibinfo {pages} {042116} (\bibinfo {year} {2019})}\BibitemShut {NoStop}%
\bibitem [{\citenamefont {{Krajnik, \v{Z}iga and Ilievski, Enej and Prosen, Toma\v{z} and Pasquier, Vincent}}(2021)}]{KrajnikLandauLifshitz}%
  \BibitemOpen
  \bibfield  {author} {\bibinfo {author} {\bibnamefont {{Krajnik, \v{Z}iga and Ilievski, Enej and Prosen, Toma\v{z} and Pasquier, Vincent}}},\ }\bibfield  {journal} {\bibinfo  {journal} {SciPost Physics}\ }\textbf {\bibinfo {volume} {11}},\ \href {https://doi.org/10.21468/scipostphys.11.3.051} {10.21468/scipostphys.11.3.051} (\bibinfo {year} {2021})\BibitemShut {NoStop}%
\bibitem [{\citenamefont {De~Nardis}\ \emph {et~al.}(2023)\citenamefont {De~Nardis}, \citenamefont {Gopalakrishnan},\ and\ \citenamefont {Vasseur}}]{NonlinearFH_JDN}%
  \BibitemOpen
  \bibfield  {author} {\bibinfo {author} {\bibfnamefont {J.}~\bibnamefont {De~Nardis}}, \bibinfo {author} {\bibfnamefont {S.}~\bibnamefont {Gopalakrishnan}},\ and\ \bibinfo {author} {\bibfnamefont {R.}~\bibnamefont {Vasseur}},\ }\href {https://doi.org/10.1103/PhysRevLett.131.197102} {\bibfield  {journal} {\bibinfo  {journal} {Phys. Rev. Lett.}\ }\textbf {\bibinfo {volume} {131}},\ \bibinfo {pages} {197102} (\bibinfo {year} {2023})}\BibitemShut {NoStop}%
\bibitem [{\citenamefont {Doyon}\ \emph {et~al.}(2023)\citenamefont {Doyon}, \citenamefont {Perfetto}, \citenamefont {Sasamoto},\ and\ \citenamefont {Yoshimura}}]{doyon-perfetto-sasamoto-yoshimura-2023}%
  \BibitemOpen
  \bibfield  {author} {\bibinfo {author} {\bibfnamefont {B.}~\bibnamefont {Doyon}}, \bibinfo {author} {\bibfnamefont {G.}~\bibnamefont {Perfetto}}, \bibinfo {author} {\bibfnamefont {T.}~\bibnamefont {Sasamoto}},\ and\ \bibinfo {author} {\bibfnamefont {T.}~\bibnamefont {Yoshimura}},\ }\href {https://doi.org/10.21468/SciPostPhys.15.4.136} {\bibfield  {journal} {\bibinfo  {journal} {SciPost Phys.}\ }\textbf {\bibinfo {volume} {15}},\ \bibinfo {pages} {136} (\bibinfo {year} {2023})}\BibitemShut {NoStop}%
\bibitem [{\citenamefont {{Zadnik, Lenart and Ljubotina, Marko and Krajnik, \v{Z}iga and Ilievski, Enej and Prosen, Toma\v{z}}}(2024)}]{Ratchets}%
  \BibitemOpen
  \bibfield  {author} {\bibinfo {author} {\bibnamefont {{Zadnik, Lenart and Ljubotina, Marko and Krajnik, \v{Z}iga and Ilievski, Enej and Prosen, Toma\v{z}}}},\ }\href {https://doi.org/10.1103/PRXQuantum.5.030356} {\bibfield  {journal} {\bibinfo  {journal} {PRX Quantum}\ }\textbf {\bibinfo {volume} {5}},\ \bibinfo {pages} {030356} (\bibinfo {year} {2024})}\BibitemShut {NoStop}%
\bibitem [{\citenamefont {Valli}\ \emph {et~al.}(2025)\citenamefont {Valli}, \citenamefont {Moca}, \citenamefont {Werner}, \citenamefont {Kormos}, \citenamefont {Krajnik}, \citenamefont {Prosen},\ and\ \citenamefont {Zar{\'a}nd}}]{valli2024efficientcomputationcumulantevolution}%
  \BibitemOpen
  \bibfield  {author} {\bibinfo {author} {\bibfnamefont {A.}~\bibnamefont {Valli}}, \bibinfo {author} {\bibfnamefont {C.~P.}\ \bibnamefont {Moca}}, \bibinfo {author} {\bibfnamefont {M.~A.}\ \bibnamefont {Werner}}, \bibinfo {author} {\bibfnamefont {M.}~\bibnamefont {Kormos}}, \bibinfo {author} {\bibfnamefont {{\v{Z}}.}~\bibnamefont {Krajnik}}, \bibinfo {author} {\bibfnamefont {T.}~\bibnamefont {Prosen}},\ and\ \bibinfo {author} {\bibfnamefont {G.}~\bibnamefont {Zar{\'a}nd}},\ }\href {https://doi.org/10.1103/f3c4-n21z} {\bibfield  {journal} {\bibinfo  {journal} {Phys. Rev. Lett.}\ }\textbf {\bibinfo {volume} {135}},\ \bibinfo {pages} {100401} (\bibinfo {year} {2025})}\BibitemShut {NoStop}%
\bibitem [{\citenamefont {Popkov}\ and\ \citenamefont {Zvyagin}(1993)}]{Popkov1993Antichiral}%
  \BibitemOpen
  \bibfield  {author} {\bibinfo {author} {\bibfnamefont {V.~{\relax Yu}.}\ \bibnamefont {Popkov}}\ and\ \bibinfo {author} {\bibfnamefont {A.~A.}\ \bibnamefont {Zvyagin}},\ }\href {https://doi.org/10.1016/0375-9601(93)90624-9} {\bibfield  {journal} {\bibinfo  {journal} {Phys. Lett. A}\ }\textbf {\bibinfo {volume} {175}},\ \bibinfo {pages} {295} (\bibinfo {year} {1993})}\BibitemShut {NoStop}%
\bibitem [{\citenamefont {Frahm}\ and\ \citenamefont {{R\"o}denbeck}(1996)}]{Frahm1996Integrable}%
  \BibitemOpen
  \bibfield  {author} {\bibinfo {author} {\bibfnamefont {H.}~\bibnamefont {Frahm}}\ and\ \bibinfo {author} {\bibfnamefont {C.}~\bibnamefont {{R\"o}denbeck}},\ }\href {https://doi.org/10.1209/epl/i1996-00302-7} {\bibfield  {journal} {\bibinfo  {journal} {Europhys. Lett.}\ }\textbf {\bibinfo {volume} {33}},\ \bibinfo {pages} {47} (\bibinfo {year} {1996})}\BibitemShut {NoStop}%
\bibitem [{\citenamefont {Frahm}\ and\ \citenamefont {{R\"o}denbeck}(1997)}]{Frahm1997Properties}%
  \BibitemOpen
  \bibfield  {author} {\bibinfo {author} {\bibfnamefont {H.}~\bibnamefont {Frahm}}\ and\ \bibinfo {author} {\bibfnamefont {C.}~\bibnamefont {{R\"o}denbeck}},\ }\href {https://doi.org/10.1088/0305-4470/30/13/005} {\bibfield  {journal} {\bibinfo  {journal} {J. Phys. A: Math. Gen.}\ }\textbf {\bibinfo {volume} {30}},\ \bibinfo {pages} {4467} (\bibinfo {year} {1997})}\BibitemShut {NoStop}%
\bibitem [{\citenamefont {Doyon}\ and\ \citenamefont {Spohn}(2017)}]{doyon2017drude}%
  \BibitemOpen
  \bibfield  {author} {\bibinfo {author} {\bibfnamefont {B.}~\bibnamefont {Doyon}}\ and\ \bibinfo {author} {\bibfnamefont {H.}~\bibnamefont {Spohn}},\ }\href@noop {} {\bibfield  {journal} {\bibinfo  {journal} {SciPost Physics}\ }\textbf {\bibinfo {volume} {3}},\ \bibinfo {pages} {039} (\bibinfo {year} {2017})}\BibitemShut {NoStop}%
\bibitem [{\citenamefont {Doyon}(2020)}]{doyon2020lecture}%
  \BibitemOpen
  \bibfield  {author} {\bibinfo {author} {\bibfnamefont {B.}~\bibnamefont {Doyon}},\ }\href@noop {} {\bibfield  {journal} {\bibinfo  {journal} {SciPost Physics Lecture Notes}\ ,\ \bibinfo {pages} {018}} (\bibinfo {year} {2020})}\BibitemShut {NoStop}%
\bibitem [{\citenamefont {Doyon}\ \emph {et~al.}(2025)\citenamefont {Doyon}, \citenamefont {Gopalakrishnan}, \citenamefont {Møller}, \citenamefont {Schmiedmayer},\ and\ \citenamefont {Vasseur}}]{Doyon_2025}%
  \BibitemOpen
  \bibfield  {author} {\bibinfo {author} {\bibfnamefont {B.}~\bibnamefont {Doyon}}, \bibinfo {author} {\bibfnamefont {S.}~\bibnamefont {Gopalakrishnan}}, \bibinfo {author} {\bibfnamefont {F.}~\bibnamefont {Møller}}, \bibinfo {author} {\bibfnamefont {J.}~\bibnamefont {Schmiedmayer}},\ and\ \bibinfo {author} {\bibfnamefont {R.}~\bibnamefont {Vasseur}},\ }\bibfield  {journal} {\bibinfo  {journal} {Physical Review X}\ }\textbf {\bibinfo {volume} {15}},\ \href {https://doi.org/10.1103/physrevx.15.010501} {10.1103/physrevx.15.010501} (\bibinfo {year} {2025})\BibitemShut {NoStop}%
\bibitem [{Note1()}]{Note1}%
  \BibitemOpen
  \bibinfo {note} {Note that our definition of $H_\chi $ differs by a factor of $1/2$ from Ref.~\cite {Frahm1997Properties}, but is consistent with the model Hamiltonian reported in Ref.~\cite {Frahm1996Integrable}, and Bethe ansatz equations found in Refs.~\cite {Popkov1993Antichiral,Frahm1996Integrable,Frahm1997Properties}.}\BibitemShut {Stop}%
\bibitem [{\citenamefont {Wen}(2002)}]{WenPSG}%
  \BibitemOpen
  \bibfield  {author} {\bibinfo {author} {\bibfnamefont {X.-G.}\ \bibnamefont {Wen}},\ }\href {https://doi.org/10.1103/PhysRevB.65.165113} {\bibfield  {journal} {\bibinfo  {journal} {Phys. Rev. B}\ }\textbf {\bibinfo {volume} {65}},\ \bibinfo {pages} {165113} (\bibinfo {year} {2002})}\BibitemShut {NoStop}%
\bibitem [{\citenamefont {Gorohovsky}\ \emph {et~al.}(2015)\citenamefont {Gorohovsky}, \citenamefont {Pereira},\ and\ \citenamefont {Sela}}]{Gorohovsky2015Chiral}%
  \BibitemOpen
  \bibfield  {author} {\bibinfo {author} {\bibfnamefont {G.}~\bibnamefont {Gorohovsky}}, \bibinfo {author} {\bibfnamefont {R.~G.}\ \bibnamefont {Pereira}},\ and\ \bibinfo {author} {\bibfnamefont {E.}~\bibnamefont {Sela}},\ }\href {https://doi.org/10.1103/PhysRevB.91.245139} {\bibfield  {journal} {\bibinfo  {journal} {Phys. Rev. B}\ }\textbf {\bibinfo {volume} {91}},\ \bibinfo {pages} {245139} (\bibinfo {year} {2015})}\BibitemShut {NoStop}%
\bibitem [{\citenamefont {Willsher}\ and\ \citenamefont {Knolle}(2025)}]{willsher2025dynamics}%
  \BibitemOpen
  \bibfield  {author} {\bibinfo {author} {\bibfnamefont {J.}~\bibnamefont {Willsher}}\ and\ \bibinfo {author} {\bibfnamefont {J.}~\bibnamefont {Knolle}},\ }\href@noop {} {\bibfield  {journal} {\bibinfo  {journal} {arXiv preprint arXiv:2503.13831}\ } (\bibinfo {year} {2025})}\BibitemShut {NoStop}%
\bibitem [{\citenamefont {Oliviero}\ \emph {et~al.}(2022)\citenamefont {Oliviero}, \citenamefont {Sobral~da Silva}, \citenamefont {Andrade},\ and\ \citenamefont {Pereira}}]{oliviero2022noncoplanar}%
  \BibitemOpen
  \bibfield  {author} {\bibinfo {author} {\bibfnamefont {F.}~\bibnamefont {Oliviero}}, \bibinfo {author} {\bibfnamefont {J.~A.}\ \bibnamefont {Sobral~da Silva}}, \bibinfo {author} {\bibfnamefont {E.}~\bibnamefont {Andrade}},\ and\ \bibinfo {author} {\bibfnamefont {R.~G.}\ \bibnamefont {Pereira}},\ }\href@noop {} {\bibfield  {journal} {\bibinfo  {journal} {SciPost Physics}\ }\textbf {\bibinfo {volume} {13}},\ \bibinfo {pages} {050} (\bibinfo {year} {2022})}\BibitemShut {NoStop}%
\bibitem [{\citenamefont {{Vanicat, Matthieu and Zadnik, Lenart and Prosen, Toma\v{z}}}(2018)}]{vanicat2018}%
  \BibitemOpen
  \bibfield  {author} {\bibinfo {author} {\bibnamefont {{Vanicat, Matthieu and Zadnik, Lenart and Prosen, Toma\v{z}}}},\ }\href {https://doi.org/10.1103/PhysRevLett.121.030606} {\bibfield  {journal} {\bibinfo  {journal} {Phys. Rev. Lett.}\ }\textbf {\bibinfo {volume} {121}},\ \bibinfo {pages} {030606} (\bibinfo {year} {2018})}\BibitemShut {NoStop}%
\bibitem [{\citenamefont {{Paletta, Chiara and Duh, Urban and Pozsgay, Bal\'azs and Zadnik, Lenart}}(2025)}]{paletta2025}%
  \BibitemOpen
  \bibfield  {author} {\bibinfo {author} {\bibnamefont {{Paletta, Chiara and Duh, Urban and Pozsgay, Bal\'azs and Zadnik, Lenart}}},\ }\href {https://doi.org/10.1088/1751-8121/ade483} {\bibfield  {journal} {\bibinfo  {journal} {Journal of Physics A: Mathematical and Theoretical}\ }\textbf {\bibinfo {volume} {58}},\ \bibinfo {pages} {275001} (\bibinfo {year} {2025})}\BibitemShut {NoStop}%
\bibitem [{\citenamefont {Takahashi}(1999)}]{takahashi1999}%
  \BibitemOpen
  \bibfield  {author} {\bibinfo {author} {\bibfnamefont {M.}~\bibnamefont {Takahashi}},\ }\href@noop {} {\emph {\bibinfo {title} {Thermodynamics of One-Dimensional Solvable Models}}}\ (\bibinfo  {publisher} {Cambridge University Press},\ \bibinfo {year} {1999})\BibitemShut {NoStop}%
\bibitem [{Note2()}]{Note2}%
  \BibitemOpen
  \bibinfo {note} {A quasiparticle traversing the distance $\ell \sim |v_s t|$ encounters $\ell $ spins that can be thought of as randomly chosen from some probability distribution with a finite variance $\sigma ^2$ and zero mean. The magnetic field felt by the quasiparticle corresponds to the mean field $h$ of the sequence of $\ell $ traversed spins. For large $\ell $ (or $t$), by the central limit theorem, $\protect \sqrt {\ell }h$ is normally distributed with zero mean and variance $\sigma ^2$. Therefore, $\langle h^2\rangle \sim |v_s t|^{-1}$.}\BibitemShut {Stop}%
\bibitem [{Note3()}]{Note3}%
  \BibitemOpen
  \bibinfo {note} {As an additional check of self-consistency, we note that for slow quasiparticles with $s\sim s_*$ the background cannot be treated as frozen. These quasiparticles move with the background, i.e., they diffuse with a time-dependent diffusion constant $s_*(t)$, since $m_2(t)\sim s_*(t) t$. Relating the length scales as $|v_{s_*}t|^2\sim s_* t$, where $|v_{s_*}|\sim s_*^{-1}$, we again find $s_*(t)\sim t^{1/3}$.}\BibitemShut {Stop}%
\bibitem [{\citenamefont {Pr\"ahofer}\ and\ \citenamefont {Spohn}(2000)}]{PrahoferSpohn}%
  \BibitemOpen
  \bibfield  {author} {\bibinfo {author} {\bibfnamefont {M.}~\bibnamefont {Pr\"ahofer}}\ and\ \bibinfo {author} {\bibfnamefont {H.}~\bibnamefont {Spohn}},\ }\href {https://doi.org/10.1103/PhysRevLett.84.4882} {\bibfield  {journal} {\bibinfo  {journal} {Phys. Rev. Lett.}\ }\textbf {\bibinfo {volume} {84}},\ \bibinfo {pages} {4882} (\bibinfo {year} {2000})}\BibitemShut {NoStop}%
\bibitem [{\citenamefont {Baik}\ and\ \citenamefont {Rains}(2000)}]{baik2000limiting}%
  \BibitemOpen
  \bibfield  {author} {\bibinfo {author} {\bibfnamefont {J.}~\bibnamefont {Baik}}\ and\ \bibinfo {author} {\bibfnamefont {E.~M.}\ \bibnamefont {Rains}},\ }\href@noop {} {\bibfield  {journal} {\bibinfo  {journal} {J. Stat. Phys.}\ }\textbf {\bibinfo {volume} {100}},\ \bibinfo {pages} {523} (\bibinfo {year} {2000})}\BibitemShut {NoStop}%
\bibitem [{\citenamefont {Tracy}\ and\ \citenamefont {Widom}(1994)}]{Tracy1994}%
  \BibitemOpen
  \bibfield  {author} {\bibinfo {author} {\bibfnamefont {C.~A.}\ \bibnamefont {Tracy}}\ and\ \bibinfo {author} {\bibfnamefont {H.}~\bibnamefont {Widom}},\ }\href {https://doi.org/10.1007/BF02100489} {\bibfield  {journal} {\bibinfo  {journal} {Communications in Mathematical Physics}\ }\textbf {\bibinfo {volume} {159}},\ \bibinfo {pages} {151} (\bibinfo {year} {1994})}\BibitemShut {NoStop}%
\bibitem [{\citenamefont {Kalinowski}\ \emph {et~al.}(2023)\citenamefont {Kalinowski}, \citenamefont {Maskara},\ and\ \citenamefont {Lukin}}]{Kalinowski2023NonAbelian}%
  \BibitemOpen
  \bibfield  {author} {\bibinfo {author} {\bibfnamefont {M.}~\bibnamefont {Kalinowski}}, \bibinfo {author} {\bibfnamefont {N.}~\bibnamefont {Maskara}},\ and\ \bibinfo {author} {\bibfnamefont {M.~D.}\ \bibnamefont {Lukin}},\ }\href {https://doi.org/10.1103/PhysRevX.13.031008} {\bibfield  {journal} {\bibinfo  {journal} {Phys. Rev. X}\ }\textbf {\bibinfo {volume} {13}},\ \bibinfo {pages} {031008} (\bibinfo {year} {2023})}\BibitemShut {NoStop}%
\bibitem [{\citenamefont {Evered}\ \emph {et~al.}(2025)\citenamefont {Evered}, \citenamefont {Kalinowski}, \citenamefont {Geim}, \citenamefont {Manovitz}, \citenamefont {Bluvstein}, \citenamefont {Li}, \citenamefont {Maskara}, \citenamefont {Zhou}, \citenamefont {Ebadi}, \citenamefont {Xu} \emph {et~al.}}]{Evered2025Probing}%
  \BibitemOpen
  \bibfield  {author} {\bibinfo {author} {\bibfnamefont {S.~J.}\ \bibnamefont {Evered}}, \bibinfo {author} {\bibfnamefont {M.}~\bibnamefont {Kalinowski}}, \bibinfo {author} {\bibfnamefont {A.~A.}\ \bibnamefont {Geim}}, \bibinfo {author} {\bibfnamefont {T.}~\bibnamefont {Manovitz}}, \bibinfo {author} {\bibfnamefont {D.}~\bibnamefont {Bluvstein}}, \bibinfo {author} {\bibfnamefont {S.~H.}\ \bibnamefont {Li}}, \bibinfo {author} {\bibfnamefont {N.}~\bibnamefont {Maskara}}, \bibinfo {author} {\bibfnamefont {H.}~\bibnamefont {Zhou}}, \bibinfo {author} {\bibfnamefont {S.}~\bibnamefont {Ebadi}}, \bibinfo {author} {\bibfnamefont {M.}~\bibnamefont {Xu}}, \emph {et~al.},\ }\href@noop {} {\bibfield  {journal} {\bibinfo  {journal} {Nature}\ }\textbf {\bibinfo {volume} {645}},\ \bibinfo {pages} {341} (\bibinfo {year} {2025})}\BibitemShut {NoStop}%
\bibitem [{Note4()}]{Note4}%
  \BibitemOpen
  \bibinfo {note} {We implement open boundary conditions by removing one of the four-site unitary gate, such that $\protect \mathbb {U}_\protect \mathrm {OBC}=\DOTSB \prod@ \slimits@ _{j=1}^{\protect \frac {L}{4}-1}U_{[4j-1,4j+2]}\DOTSB \prod@ \slimits@ _{j=1}^{\protect \frac {L}{4}}U_{[4j-3,4j]}$}\BibitemShut {NoStop}%
\bibitem [{\citenamefont {Fishman}\ \emph {et~al.}(2022{\natexlab{a}})\citenamefont {Fishman}, \citenamefont {White},\ and\ \citenamefont {Stoudenmire}}]{Matthew2022itensor}%
  \BibitemOpen
  \bibfield  {author} {\bibinfo {author} {\bibfnamefont {M.}~\bibnamefont {Fishman}}, \bibinfo {author} {\bibfnamefont {S.~R.}\ \bibnamefont {White}},\ and\ \bibinfo {author} {\bibfnamefont {E.~M.}\ \bibnamefont {Stoudenmire}},\ }\href {https://doi.org/10.21468/SciPostPhysCodeb.4} {\bibfield  {journal} {\bibinfo  {journal} {SciPost Phys. Codebases}\ ,\ \bibinfo {pages} {4}} (\bibinfo {year} {2022}{\natexlab{a}})}\BibitemShut {NoStop}%
\bibitem [{\citenamefont {Fishman}\ \emph {et~al.}(2022{\natexlab{b}})\citenamefont {Fishman}, \citenamefont {White},\ and\ \citenamefont {Stoudenmire}}]{Matthew2022codebase}%
  \BibitemOpen
  \bibfield  {author} {\bibinfo {author} {\bibfnamefont {M.}~\bibnamefont {Fishman}}, \bibinfo {author} {\bibfnamefont {S.~R.}\ \bibnamefont {White}},\ and\ \bibinfo {author} {\bibfnamefont {E.~M.}\ \bibnamefont {Stoudenmire}},\ }\href {https://doi.org/10.21468/SciPostPhysCodeb.4-r0.3} {\bibfield  {journal} {\bibinfo  {journal} {SciPost Phys. Codebases}\ ,\ \bibinfo {pages} {4}} (\bibinfo {year} {2022}{\natexlab{b}})}\BibitemShut {NoStop}%
\bibitem [{\citenamefont {Weinberg}\ and\ \citenamefont {Bukov}(2017)}]{philip2017quspin}%
  \BibitemOpen
  \bibfield  {author} {\bibinfo {author} {\bibfnamefont {P.}~\bibnamefont {Weinberg}}\ and\ \bibinfo {author} {\bibfnamefont {M.}~\bibnamefont {Bukov}},\ }\href {https://doi.org/10.21468/SciPostPhys.2.1.003} {\bibfield  {journal} {\bibinfo  {journal} {SciPost Phys.}\ }\textbf {\bibinfo {volume} {2}},\ \bibinfo {pages} {003} (\bibinfo {year} {2017})}\BibitemShut {NoStop}%
\bibitem [{\citenamefont {Weinberg}\ and\ \citenamefont {Bukov}(2019)}]{philipe2019quspin}%
  \BibitemOpen
  \bibfield  {author} {\bibinfo {author} {\bibfnamefont {P.}~\bibnamefont {Weinberg}}\ and\ \bibinfo {author} {\bibfnamefont {M.}~\bibnamefont {Bukov}},\ }\href {https://doi.org/10.21468/SciPostPhys.7.2.020} {\bibfield  {journal} {\bibinfo  {journal} {SciPost Phys.}\ }\textbf {\bibinfo {volume} {7}},\ \bibinfo {pages} {020} (\bibinfo {year} {2019})}\BibitemShut {NoStop}%
\bibitem [{\citenamefont {Faddeev}(1996)}]{faddeev1996}%
  \BibitemOpen
  \bibfield  {author} {\bibinfo {author} {\bibfnamefont {L.~D.}\ \bibnamefont {Faddeev}},\ }\href {https://arxiv.org/abs/hep-th/9605187} {\bibinfo {title} {{How Algebraic Bethe Ansatz works for integrable model}}} (\bibinfo {year} {1996}),\ \Eprint {https://arxiv.org/abs/hep-th/9605187} {arXiv:hep-th/9605187 [hep-th]} \BibitemShut {NoStop}%
\end{thebibliography}%

\end{document}